\documentclass{article}

\usepackage{appendix}

\usepackage{amsmath}
\usepackage{amsfonts}
\usepackage{amssymb}

\usepackage[style=chem-acs,maxbibnames=99]{biblatex}
\bibliography{citations/Paperpile_LocalMD_Jul_06}

\usepackage{graphicx}

\usepackage{url}

\usepackage{algorithm}
\usepackage{algpseudocode}

\usepackage{geometry}
\usepackage{authblk}

\newcommand{\xs}{\mathbf{x}}
\newcommand{\zs}{\mathbf{z}}

\title{A local resampling trick\\for focused molecular dynamics}
\author[1,$\dag$]{Joshua Fass}
\author[1,$\dag$]{Forrest York}
\author[1]{Matthew Wittmann}
\author[1]{Joseph Kaus}
\author[1]{Yutong Zhao*}
\affil[1]{Computation, Relay Therapeutics, Cambridge, MA, 02139, US}
\affil[*]{Correspondence: \texttt{yzhao@relaytx.com}}
\date{August 18, 2023}

\begin{document}
\newrefsection

\maketitle

\begin{abstract}
We describe a method that focuses sampling effort on a user-defined selection of a large system, which can lead to substantial decreases in computational effort by speeding up the calculation of nonbonded interactions. 
A naive approach can lead to incorrect sampling if the selection depends on the configuration in a way that is not accounted for.
We avoid this pitfall by introducing appropriate auxiliary variables.
This results in an implementation that is closely related to ``configurational freezing'' and ``elastic barrier dynamical freezing.''
We implement the method and validate that it can be used to supplement conventional molecular dynamics in free energy calculations (absolute hydration and relative binding).
\end{abstract}

\section{Introduction}

\subsection{Motivation}
In molecular simulation, it is often useful to focus sampling effort on small parts of a large system.
Examples may include:
(1) cases where a small perturbation is introduced to a large system, such as relative binding free energy calculations in explicit solvent,
(2) cases where a more efficient sampler can be implemented for a subsystem than for the system overall, and
(3) cases where the observable being computed depends most sensitively on fluctuations in a small region of the system.

To focus effort on a small region, it may be tempting to select particles in an arbitrary spatial region of the system, allowing the selected particles to move freely while the rest are held fixed.
However, this approach can lead to incorrect sampling if the selection of each particle depends on the system's configuration in a way that is not accounted for.

\begin{figure}
    \centering
    \includegraphics[width=\columnwidth]{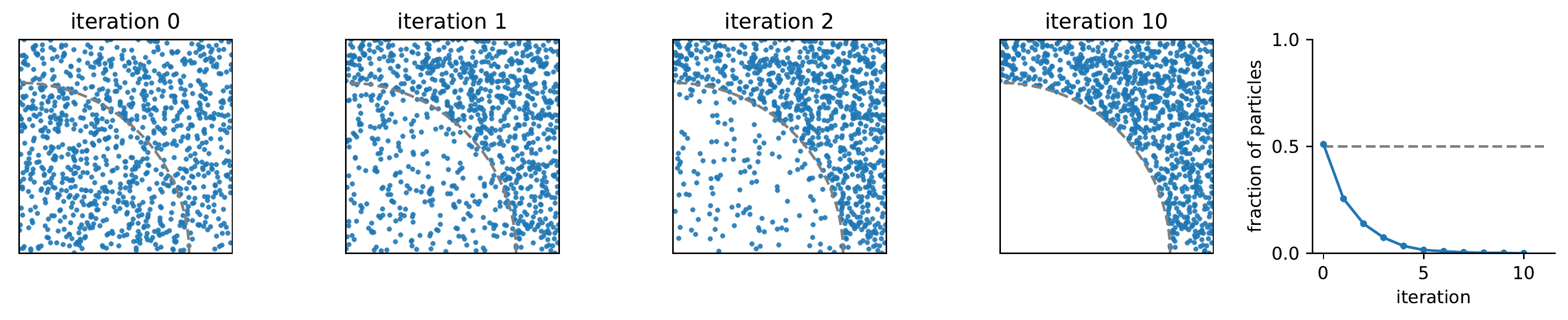}
    \caption{\textbf{Naive conditional resampling can lead to dramatic sampling biases.}
    In the case of non-interacting particles, selecting particles for resampling only if they are in some region will eventually remove all particles from that region.
    The first 4 panels show the configuration of the system as resampling moves are repeatedly applied to a region.
    The last panel shows the fraction of particles within that region.
    A similar observation was made in \cite{Giovannelli2016-fx}.
    }    \label{fig:naive_conditional_resampling}
\end{figure}

This pitfall can be illustrated using a box of non-interacting particles.
Say we are most interested in the region of this system within a distance $r$ of the origin.
At equilibrium, particles are distributed uniformly through the box.
If we repeatedly select all particles $i$ where $\|x_i\| < r$, resampling the positions of these particles while freezing the rest, then eventually we will remove all particles from the region of interest, as illustrated in Figure \ref{fig:naive_conditional_resampling}.
A similar observation was made by Giovannelli et al. in \cite{Giovannelli2016-fx}.

\subsection{Outline}
In Section \ref{section:local_resampling}, we suggest an auxiliary variable method to account for user-defined particle selection probabilities, and consider a series of specializations that may be convenient to implement in MD.
We then investigate implementations and applications of this method, in the context of absolute hydration free energies (AHFE) and relative binding free energies (RBFE), in Section \ref{section:applications}.
We discuss limitations, possibilities for future work, and the remaining gap between expected and realized speed-up in Section \ref{section:discussion}.
We discuss related work in Appendix \ref{appendix:related_work}.

\section{Local resampling using auxiliary variables}
\label{section:local_resampling}

We review auxiliary variable methods in Section \ref{section:aux_variable_methods}, then use auxiliary variables to account for 
user-defined selection probabilities in Section \ref{section:local_resampling_recipe}.
We then consider a sequence of specializations that lead to a convenient MD implementation: first restricting the selection probabilities to be independent in Section \ref{section:independent_selection}, then restricting these probabilities to depend only on distance to a distinguished particle in Section \ref{section:radial_selection}.
This leads to an implementation that involves adding a simple radial flat-bottom restraint term to the system's potential energy function, but assumes access to exact MCMC moves.
In Section \ref{subsection:intro_local_md} we discuss implementing local resampling using an approximate MCMC move, Langevin MD, and conclude with Section \ref{subsection:local_with_unadjusted} which assesses local resampling for bias using Langevin MD.

\subsection{Background: auxiliary variable methods}
\label{section:aux_variable_methods}

Here we review auxiliary variable methods in a general context, following \cite{Higdon1998-hz}.

Given a sampling problem $p_\text{target}(\xs)$, we can introduce any auxiliary variable $\zs$ using any conditional distribution $p_\text{auxiliary}(\zs | \xs)$.
Since the joint distribution of $\xs$ and $\zs$ is $p_\text{joint}(\xs, \zs) = p_\text{target}(\xs) \cdot p_\text{auxiliary}(\zs | \xs)$, the $\xs$-marginal of $p_\text{joint}$ is equal to $p_\text{target}$, for any choice of $p_\text{auxiliary}$.
Note that $\zs$ does not have to be defined in the same space as $\xs$.
A sampler can perform updates that involve augmenting $\xs$ with $\zs | \xs$, updating $\xs$ (and optionally $\zs$) in a way that preserves $p_\text{joint}(\xs, \zs)$, and then discarding $\zs$.
This is summarized in Algorithm \ref{alg:aux_var_sampling}.

\begin{algorithm}
\caption{Auxiliary variable sampling}
\label{alg:aux_var_sampling}
\begin{algorithmic}
\Require current state $\xs_0$
\Require target distribution $p_\text{target}(\xs)$
\Require conditional distribution $p_\text{auxiliary}(\zs | \xs)$
\Require transition kernel $T((\xs_1, \zs_1) | (\xs_0, \zs_0))$\\ \Comment{$T$ preserves the distribution $p_\text{joint}(\xs, \zs) = p_\text{target}(\xs) \cdot p_\text{auxiliary}(\zs | \xs)$}

\\
\State sample $\zs_0 \sim p_\text{auxiliary}(\cdot | \xs_0)$
\State sample $(\xs_1, \zs_1) \sim T(\cdot | (\xs_0, \zs_0))$
\State return $\xs_1$  \Comment{discard auxiliary variables $\zs_0$, $\zs_1$}

\end{algorithmic}
\end{algorithm}

Many familiar sampling algorithms can be seen as special cases of Algorithm \ref{alg:aux_var_sampling}, notably including Hamiltonian Monte Carlo (HMC)~\cite{Neal2012-ya} (which augments the configuration variables $\xs$ with velocity variables $\mathbf{v}$ and simulates dynamics that preserve $p(\xs, \mathbf{v})$).

\subsection{Generic local resampling recipe}
\label{section:local_resampling_recipe}

Define the auxiliary variable $\zs$ to be a vector of binary variables conditioned on $\xs$: $p_\text{selection}(\zs | \xs)$.
Each variable $z_i$ will represent whether particle $i$ is ``selected''.
A generic sampler that only updates the positions of the selected particles is summarized in Algorithm \ref{alg:generic_local_resampling}.

\begin{algorithm}
\caption{Local resampling}
\label{alg:generic_local_resampling}
\begin{algorithmic}
\Require current state $\xs_0$ \Comment{$\xs_0 \in \mathcal{X}^N$}
\Require $q_\text{target}(\xs)$ \Comment{defined on $\mathcal{X}^N$}
\Require $p_\text{selection}(\zs | \xs)$ \Comment{$\zs \in \left\{ 0, 1 \right\}^N$}
\Require transition kernel $T_q(\xs^\text{sub}_1 | \xs^\text{sub}_0)$
\Comment{preserves any target $q$ defined on $\mathcal{X}^K$, $K \leq N$}

\\
\State sample $\zs_0 \sim p_\text{selection}(\cdot | \xs_0)$ \Comment{select $K = \texttt{sum}(\zs_0) \leq N$ particles}
\State define $q_\text{restrained}(\xs) = q_\text{target}(\xs) \cdot p_\text{selection}(\zs_0 | \xs)$ \Comment{defined on $\mathcal{X}^N$}
\State set $\xs_0^\text{sub} \gets \xs_0[\zs_0]$
\Comment{$\xs_0^\text{sub} \in \mathcal{X}^K$}
\State define $\texttt{expand}(\xs^\text{sub}) = \xs_0.\texttt{at}[\zs_0].\texttt{set}(\xs^\text{sub})$
\Comment{set positions of selected particles}
\State define $q(\xs^\text{sub}) = q_\text{restrained}(\texttt{expand}(\xs^\text{sub}))$ \Comment{defined on $\mathcal{X}^K$}
\State sample $\xs^{\text{sub}}_1 \sim T_q(\cdot | \xs_0^\text{sub})$ \Comment{$\xs^{\text{sub}}_1 \in \mathcal{X}^K$}
\State set $\xs_1 \gets \texttt{expand}(\xs^\text{sub}_1)$ \Comment{$\xs_1 \in \mathcal{X}^N$}
\State return $\xs_1$  \Comment{discard auxiliary selection variables $\zs_0$}
\end{algorithmic}
\end{algorithm}


Intuitively, this method samples a particle selection conditioned on the current state, defines a restrained subproblem on the selected particles, applies a valid move for that restrained subproblem, and then discards the variables representing the selection.
A valid move $T_q$ is a transition kernel from which we can sample, $x_1 \sim T_q(\cdot | x_0)$, and which has $q$ as an invariant distribution (if $x_t$ is distributed according to $q$, then $x_{t+1} \sim T_q(\cdot | x_t)$ is distributed according to $q$).
Assuming the transition kernel $T_q$ is a valid $q$-invariant move (such as an MCMC move targeting $q$), then Algorithm \ref{alg:generic_local_resampling} is a valid $q_\text{target}$-invariant move, for any choice of $p_\text{selection}$, since it is a special case of Algorithm \ref{alg:aux_var_sampling}.

Next we will elaborate on specific choices of selection function $p_\text{selection}$ and transition kernel $T_q$ that may permit efficient implementation in the context of molecular simulation.

\subsection{Independent particle selection}
\label{section:independent_selection}

It will be convenient to restrict attention to $p_\text{selection}$ functions where the probabilities of selecting each particle $i$ are independent.
In other words,
\begin{align}
    z_i \mid \xs &\sim \text{Bernoulli}(f_i(\xs))\\
    p_\text{selection}(\zs \mid \xs)
    &= \prod_i \left( z_i f_i(\xs) + (1 - z_i) (1 - f_i(\xs)) \right) \\
    &= \prod_{i; z_i = 1} f_i(\xs) \prod_{i; z_i = 0} (1 - f_i(\xs))  \label{eq:restraint}
\end{align}
where each $f_i(\xs)$ is an arbitrary user-defined function $f_i : \mathcal{X}^N \to [0,1]$.
For example, we may define $f_i(\xs)$ to be close to 1 when $x_i$ is in an interesting region, and close to 0 otherwise.

This restriction has the benefits of (1) simplifying the computation of the restraint term associated with a fixed selection $\zs$, and (2) providing an intuitive user interface whose inputs are easy to validate.
For example, to assert that $p_\text{selection}(\zs | \xs)$ is normalized, it suffices to assert that the outputs of each user-defined function $f_i$ are always in the interval $[0, 1]$.
However, this restriction excludes some potentially useful choices of $p_\text{selection}$.
For example, it rules out $p_\text{selection}$ functions that could guarantee a selection of a fixed \emph{number} of particles.

\subsection{Radial selection probability}
\label{section:radial_selection}

Next we will consider a further specialization of Section \ref{section:independent_selection} intended to be convenient for implementation in molecular dynamics.
The details in this section are more likely to require problem-specific or implementation-specific adaptation than the preceding sections, and some alternative directions are considered in Section \ref{section:better_selection}.

\begin{figure}
    \centering
    \includegraphics[width=0.3 \columnwidth]{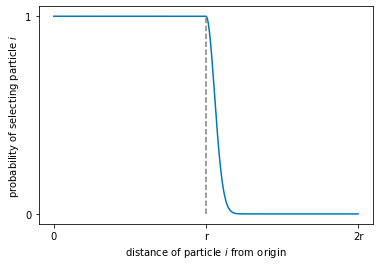}
    \includegraphics[width=0.3 \columnwidth]{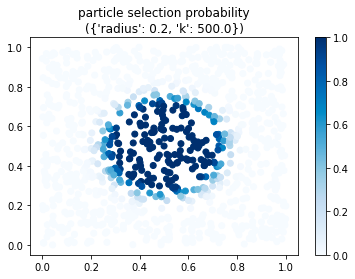}
    \includegraphics[width=0.3 \columnwidth]{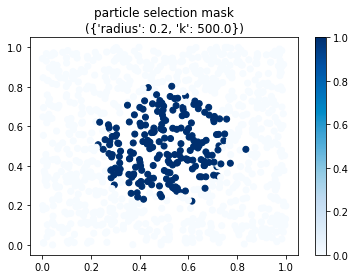}
    \caption{\textbf{Example of radial selection probability.}
    Particles $i$ within $r_0$ of the reference particle $c$ are selected with unit probability, and some further particles are also randomly selected.
    First panel: $h(r) = \exp(-\beta U_\text{fb}(r))$, second panel: particles colored by $f_i(\xs) = h(\| x_c - x_i \|)$, third panel: a random draw from $p_\text{selection}(\zs | \xs)$.\\
    Note that this illustration shows a flat-bottom \emph{harmonic} restraint ($U_\text{fb}(r) = \mathbb{I}(r > r_0) \times \frac{1}{2} k (r - r_0)^2$), but all subsequent numerical experiments use a flat-bottom \emph{quartic} restraint.
    }
    \label{fig:radial_selection}
\end{figure}

Consider a special case of Section \ref{section:independent_selection} where the functions $f_i$ each depend on a single distance involving particle $i$: $f_i(\xs) = h(r_{c,i})$, where $r_{c,i}$ is the distance between $x_i$ and a central particle $x_c$.
Define $h(r) = \exp(-\beta U_\text{fb}(r))$ where $U_\text{fb}(r; r_0, k)$ is taken to be a ``flat-bottom'' restraint potential energy function (such that $h(r) = 1$ when $r \leq r_0$, and $h(r) \to 0$ when $r \gg r_0$), and where $\beta=\frac{1}{k_B T}$ is defined by the temperature.
Intuitively, this choice will cause all particles within distance $r_0$ of particle $c$ to be selected with unit probability, and some particles further than $r_0$ to be selected as well, illustrated in Figure \ref{fig:radial_selection}.

Translating this into a potential energy function (by substituting into equation \ref{eq:restraint}, and applying $p(x) \propto \exp(-\beta U(x))$) will result in a sum of two groups of terms.
\begin{align}
p_\text{selection}(\zs | \xs)
&= \left( \prod_{i; z_i = 1} \exp(-\beta U_\text{fb}(r_{c, i})) \right) \times \left( \prod_{i; z_i = 0} (1 - \exp(-\beta U_\text{fb}(r_{c, i}))) \right)\\
U_\text{selection}(\xs; \zs)
&= \left( \sum_{i; z_i = 1} U_\text{fb}(r_{c,i}) \right) + \left( - \beta^{-1} \sum_{i; z_i = 0} \log \left(1 - \exp\left(-\beta U_\text{fb}(r_{c,i}) \right)\right) \right)  \label{eq:fb_restraint}
\end{align}

The effect of the first group of terms (involving the selected particles $z_i = 1$) is intuitive: keep all of the selected particles $i$ near the central particle $c$.
The second group of terms (involving the deselected particles $z_i = 0$) is less intuitive, preventing the central particle $c$ from moving in such a way that deselected particles have a much higher chance of becoming selected.

To avoid having to compute the second group  of terms, we can simply prevent the transition kernel $T$ from updating the central particle $c$ (so that applying $T$ cannot change the contributions associated with the pairs $c,i$ where $z_i=0$), and use the simplified target in eq. \ref{eq:fb_restraint_frozen_c}.

\begin{align}
U_\text{restraint}(\xs; \zs)
&= \sum_{i; z_i = 1} U_\text{fb}(r_{c,i}) + \text{constant}
\label{eq:fb_restraint_frozen_c}
\end{align}

The restraint is now in a safe and convenient form for molecular dynamics, and we used this form for all numerical experiments.
The choice to freeze the central particle is considered further in Appendix \ref{section:freezing_central_particle}.

In summary, this specialization chooses a central particle index $c$, randomly selects mobile particles with a probability that depends only on their distance to the central particle, freezes the central particle, and then simulates only the selected particles, subject to a restraint.
This specialization still exposes several free parameters: the choice of central particle index $c$, and the restraint parameters $r_0$, $k$.

The choice of central particle index $c$ can be arbitrary, as long as it does not depend on $\xs$.
A reasonable choice may be to sample $c$ uniformly at the start of each move from a predefined subset of interesting particles.
In the context of a hydration free energy calculation, this subset may include all of the solute atoms.
In the context of a 
binding free energy calculation, efficiency may be improved by including binding pocket residues in this subset.

The restraint parameters $r_0$, $k$ can be chosen to balance several considerations depending on the context, such as MD stability (very high values of $k$ may restrict the maximum stable timestep or introduce significant timestep error), tuning the size of the selected region, and reducing variability in the number of selected particles.
These parameters can also be randomized: each local resampling move can pick a different radius, force constant, and central particle index.
We will investigate the effect of the restraint radius $r_0$ and $k$ in a grid search in Section \ref{subsection:local_with_unadjusted}.

The functional form of $U_\text{fb}$ can also be tailored depending on context.
Throughout, we used a flat-bottom quartic restraint ($U_\text{quartic}(r; r_0, k) = \mathbb{I}(r > r_0) \times \frac{1}{4} k (r - r_0)^4$) to ensure that the restraint is at least $C^2$ continuous.
However, this specific choice also has the side effect of including appreciable probability mass up to some effective radius $r_\text{eff} > r_0$.
It may be beneficial to consider using barrier functions~\cite{Kolossvary2022-kj} (which guarantee $h(r) = 0$ whenever $r > r_0$) instead of restraint functions (where $h(r) > 0$ for some $r > r_0$), but we did not investigate this.

\subsection{Local Unadjusted MD Moves}
\label{subsection:intro_local_md}
The previous Section \ref{section:radial_selection} describes an MD compatible method to focus sampling on specific regions of a system using local moves.
Next, we apply two optimizations that can lead to unsound sampling in principle, but that we expect to be useful in practice: using unadjusted Langevin MD to sample the local subproblem, and alternating these local unadjusted MD (local MD) moves with all-atom unadjusted MD (global MD) moves.
These global MD moves contain MC barostat moves as well, while local MD does not.
MC barostat moves require changing the particle selections, which for simplicity and speed we disabled during local MD.
We follow the protocol described in Algorithm \ref{alg:local_md} for alternating between global and local MD.

\begin{algorithm}
\caption{Alternating Global and Local MD}
\label{alg:local_md}
\begin{algorithmic}
\Require \texttt{particle\_set} \Comment{possible choices of central particle $c$}
\For{$n$ in \texttt{range}($N_\text{samples}$)}
\State run $N_\text{global}$ steps of global MD.
\State sample central particle $c$ uniformly from \texttt{particle\_set}
\State probabilistically select particles based on distance to central particle $c$ \Comment{Section \ref{section:radial_selection}}
\State restrain selected particles to central particle $c$ using quartic restraint \Comment{using eq. \ref{eq:fb_restraint_frozen_c}}
\State freeze non-selected particles
\State run $N_\text{local}$ steps of local MD
\State store coordinates
\EndFor
\end{algorithmic}
\end{algorithm}

\begin{figure}
    \centering
    \includegraphics[width=0.9\columnwidth]{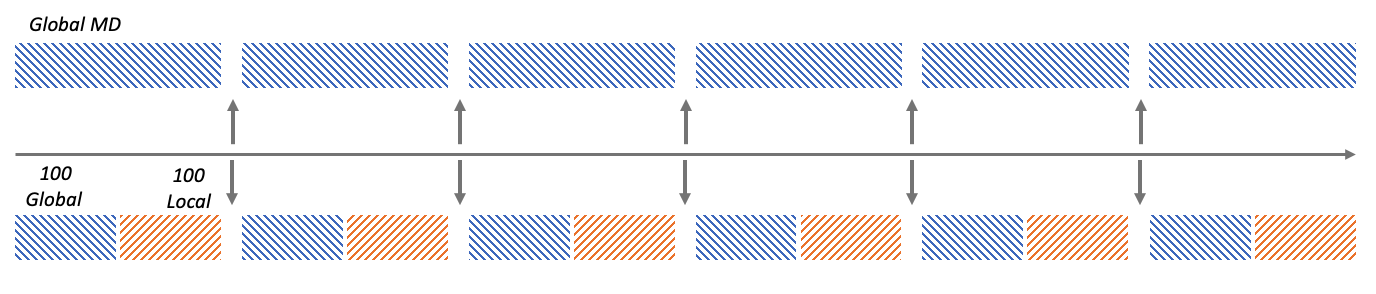}
    \caption{\textbf{Sampling alternating global and local MD moves.}
    An example, comparing a global MD protocol against alternating 100 global and 100 local MD steps between each sample.
    Each arrow indicates a sample being collected in the protocol.
    At the start of each local MD block, a new central particle is randomly selected, which is used to determine the free particles within the system.
    }\label{fig:local_md_protocol}
\end{figure}

This protocol is intended to accelerate local sampling while still retaining some ergodic global moves.
However, by using unadjusted Langevin MD and alternating between global and local moves, different free parameters (discussed in Section \ref{section:radial_selection}) may result in biased or unstable simulations.
We summarize the difference between running global MD and the protocol that mixes global and local MD in Figure \ref{fig:local_md_protocol}.
Alternating between global and local MD was chosen for simplicity, one could easily mix global and local MD in more sophisticated ways such as randomly selecting $N_\text{global}$ and $N_\text{local}$ steps.
In the next section, we evaluate different combinations of free parameters for bias and instability when alternating between local and global MD.

\subsection{Measuring Sampling Bias of Alternating Global and Local MD}
\label{subsection:local_with_unadjusted}
The previous Section \ref{subsection:intro_local_md} describes alternating between global and local MD using unadjusted Langevin MD which could result in invalid sampling.
For example, if the local and global moves are alternated sufficiently infrequently, we would expect the sampler to have very low bias (due to the use of a specific discretization of Langevin dynamics that introduces small sampling bias at steady-state~\cite{Leimkuhler2013-iz}).
Depending on how frequently these moves are alternated, the sampling can become more biased or even unstable, so we attempt to measure the extent of the bias for specific settings.

We want to determine the bias of alternating between global and local MD moves using different free parameters to determine which parameters are safe for use in practice.
We perform this filtering by computing thresholds on simulations features using samples collected with only global MD moves as the ground truth.
Once we have thresholds, we can filter parameters into acceptably biased and unacceptably biased. 

To evaluate bias, we arbitrarily selected \texttt{mobley\_9979854} from the FreeSolv set (\url{https://github.com/MobleyLab/FreeSolv/releases/tag/v0.52}). Which we ran simulations of in a 4nm TIP3P water box under different combinations of global and local MD parameters.
For complete details on the simulation setup, refer to Appendix \ref{appendix:bias_system_setup}.

We perform a grid search over parameters for global MD simulations (friction, timestep) (Table \ref{table:global_params}) as well as over parameters for simulations alternating between global and local MD (number of local steps, number of global steps, radius, $k$) (Table \ref{table:local_params}), simulating each set of parameters in triplicate with different seeds.
For each simulation, we collected the per atom ligand-environment interaction energies. These energies were used as features to perform statistical comparisons between simulations, allowing us to evaluate the bias introduced by including local MD moves.
The simulations that contained no local moves were used as the ground truth for determining the threshold of acceptable bias.

\begin{table}[h!]
\centering
\begin{tabular}{ |c|c| }
\hline
 Parameter & Values \\
 \hline\hline
 timestep (fs) & 1.5, 2.5 \\
 friction (1/ps) & 1.0, inf \\  
 \hline
\end{tabular}
\caption{Global MD parameters that were evaluated in the grid search.}
\label{table:global_params}
\end{table}

\begin{table}[h!]
\centering
\begin{tabular}{ |c|c| }
\hline
 Parameter & Values \\
 \hline\hline
 local steps & 1, 100, 300, 500 \\
 global steps & 0, 1, 10, 100, 250, 500 \\ 
 radius (nm) & 0.1, 0.2, 0.5, 1.0 \\
 $k$ (kJ/mol/nm$^4$) & 1.0, 1000.0, 10000.0 \\
 \hline
\end{tabular}
\caption{Local MD specific parameters that were evaluated in the grid search.}
\label{table:local_params}
\end{table}

\begin{figure}
    \centering
    \includegraphics[width=0.45\columnwidth]{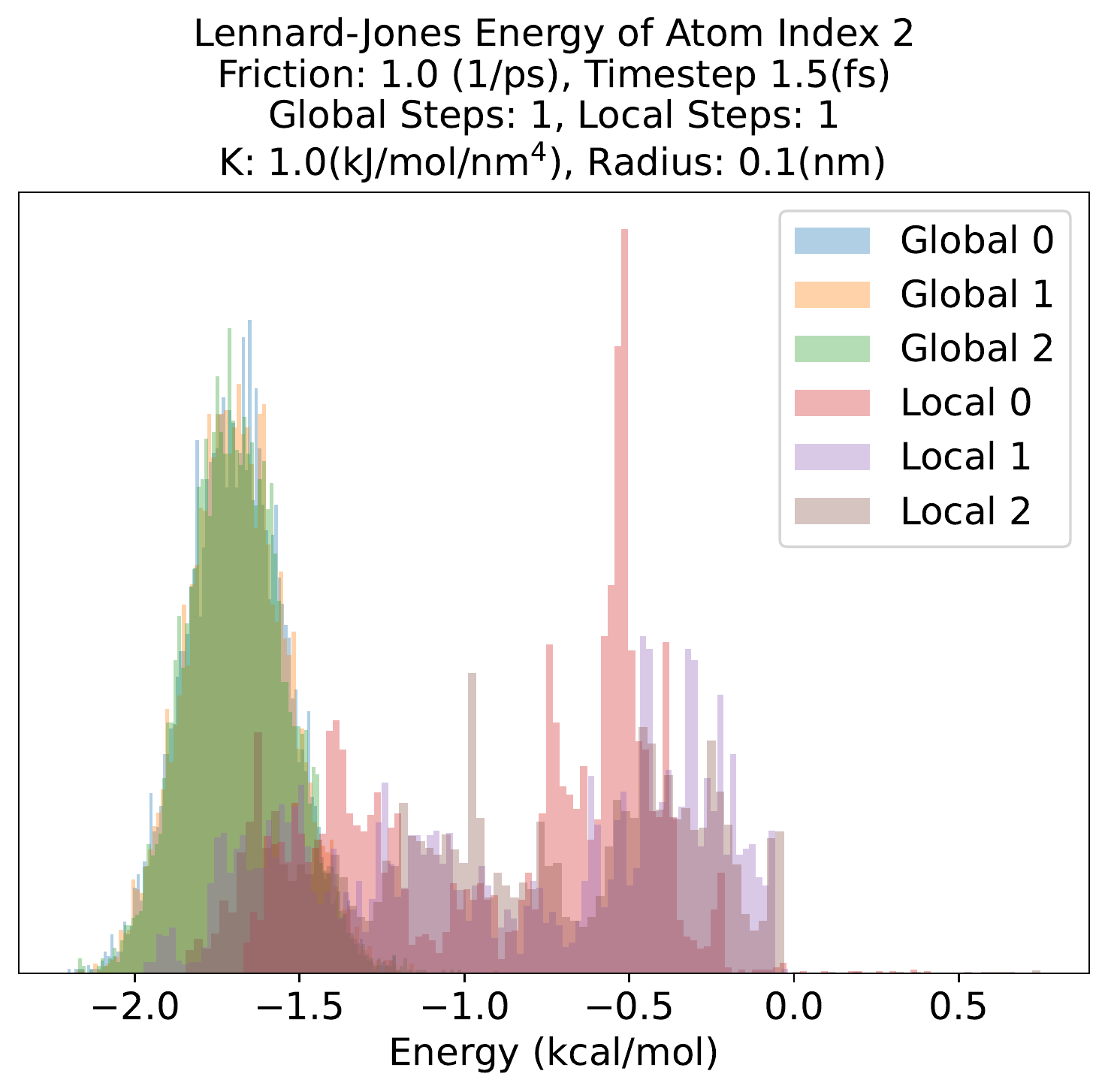}
    \includegraphics[width=0.45\columnwidth]{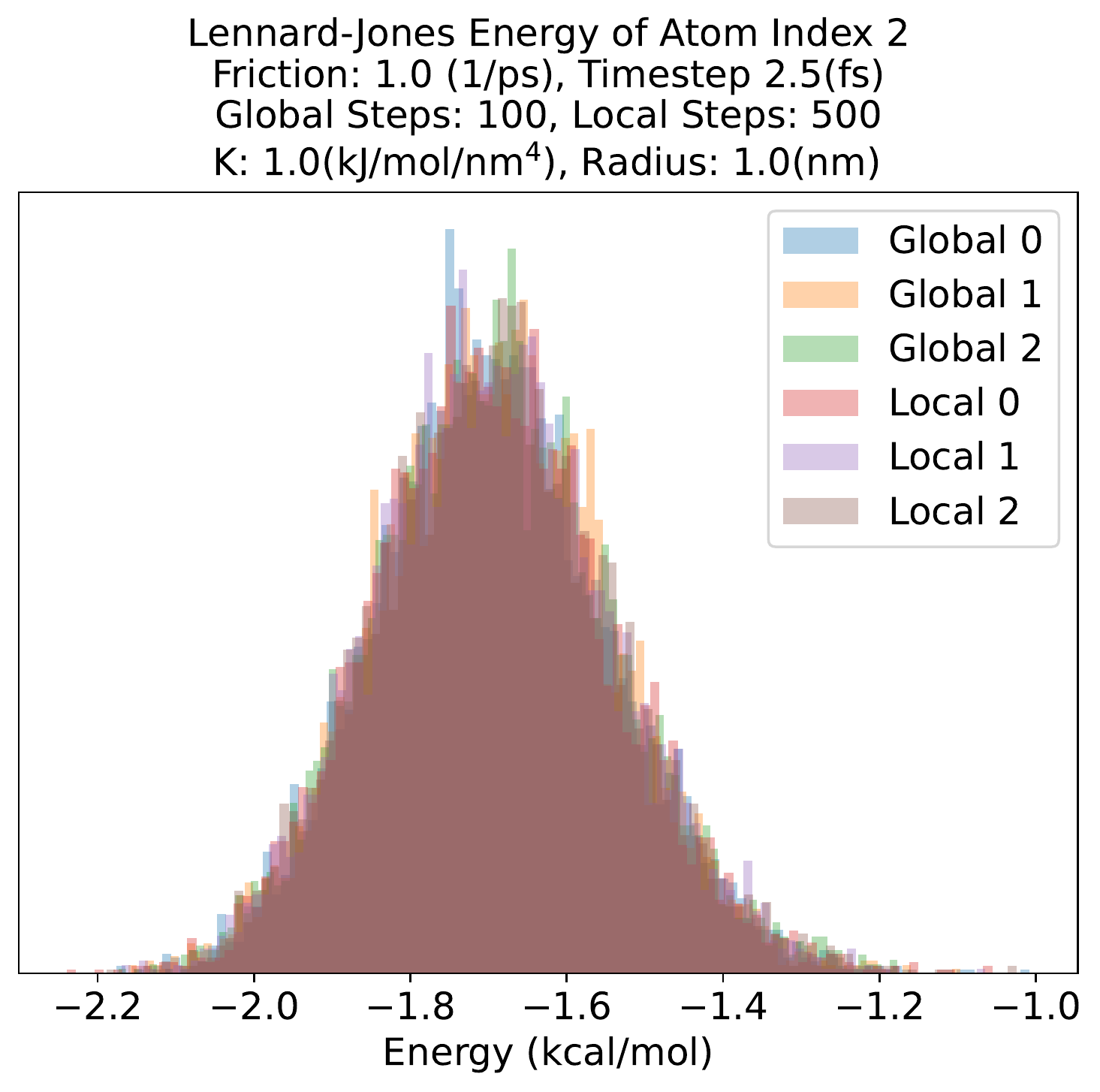}
    \caption{\textbf{Energy distributions under different local parameters.}
    First panel: Replicates of a set of parameters that introduces a large bias to the Lennard-Jones energy of an arbitrary atom in the ligand.
    Second panel: A set of parameters for alternating between global and local MD on the same atom, which shows no discernible bias in the Lennard-Jones energy of the same atom.
    }\label{fig:filtering_examples}
\end{figure}

\begin{figure}
    \centering
    \includegraphics[width=0.45\columnwidth]{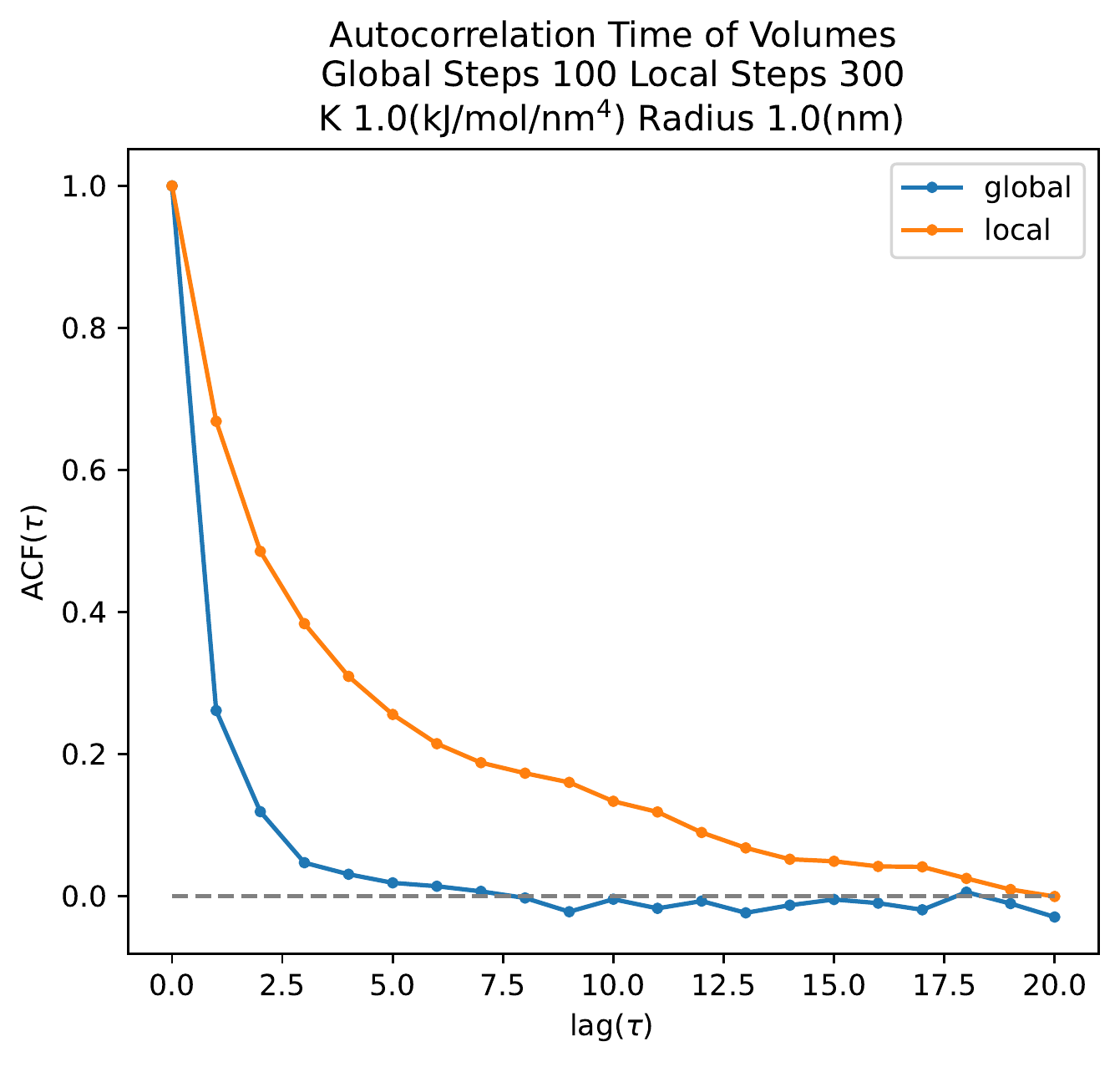}
    \includegraphics[width=0.45\columnwidth]{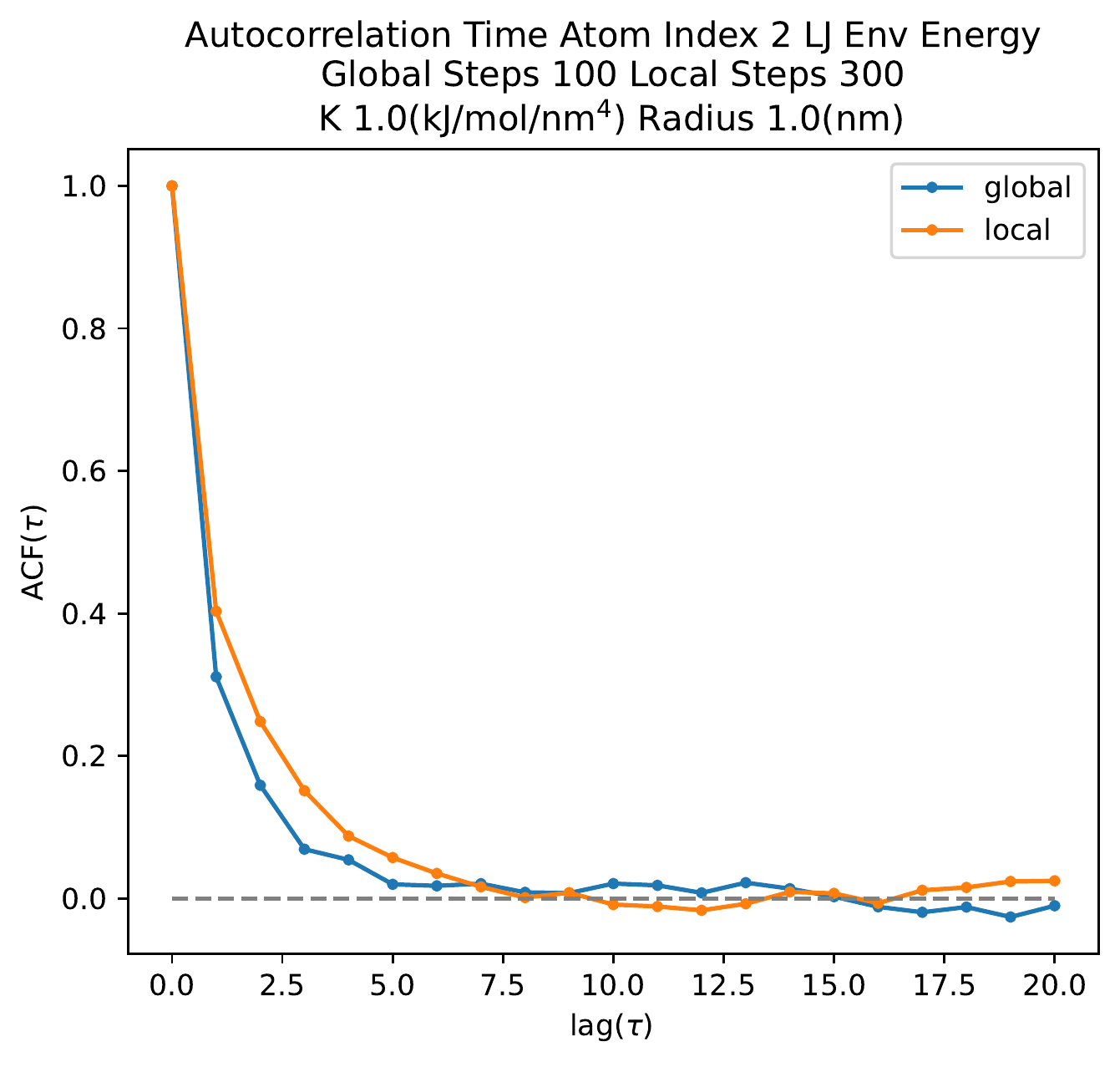}
    \caption{\textbf{Autocorrelation times of different properties may have different sensitivity to substituting local MD for global MD.}
    First panel: The autocorrelation time for box volumes with global MD only and alternating global and local MD.
    The global MD box volumes decorrelate much faster than the samples that alternate global and local MD due to disabling the MC barostat during local steps.
    With parts of the system frozen, properties that depend on frozen regions may be slower to decorrelate.
    Second panel: The autocorrelation time for an atom's Lennard-Jones environment energy with global MD and alternating global and local MD.
    The similar time for per atom energies to decorrelate under global MD and the alternating scheme, demonstrates little sensitivity to substituting global MD for local MD.
    Each lag on the x-axis represents 250 frames = 250 frames $*$ 400 steps $*$ 1.5(fs) = 0.15 nanoseconds.
    }\label{fig:convergence_difference}
\end{figure}

We use the Two-sample Kolmogorov–Smirnov test (KS test) statistic to compute per atom thresholds with the global MD samples.
These thresholds are computed using the KS Test statistic on the pairwise set of atom energies between replicates of global MD samples for each set of global MD parameters.
The samples that alternate global and local MD had a KS test statistic computed pairwise between their atom energies and that of the matching global settings (friction, timestep).
Using this filtering protocol, we found approximately 67\% of the parameters that mixed local and global MD to have acceptable bias.
Figure \ref{fig:filtering_examples} shows the difference between parameters that result in unacceptable and acceptable bias, for complete details on the filtering protocol refer to Appendix \ref{appendix:parameter_filtering}.
It is important to note that the parameters that have acceptable bias are still not ergodic and may slow converge of some system properties, as summarized in Figure \ref{fig:convergence_difference}.

From this analysis, we adopted a general rule of thumb to use global steps $\geq 100$ and local steps $\geq 100$ when alternating global and local MD.
The other parameters of radius and  $k$ searched have less significant impact, with conditions closer to global MD performing slightly better (i.e. more selected particles).
These parameters are discussed in more detail later in Section \ref{section:rbfe_only_local}.

\section{Applications}
\label{section:applications}

We first apply the unadjusted MD variant in the context of hydration free energies, using both staged calculations (Section \ref{section:staged_ahfe}), and sequential Monte Carlo (Section \ref{section:adaptive_smc}) to demonstrate the applicability of local moves to estimation of free energies.
Then, we demonstrate applicability to staged relative binding free energies in \ref{section:staged_rbfe}.


\subsection{Absolute hydration free energy}
\label{section:ahfe}

We investigated absolute hydration free energy (AHFE) predictions using local MD with a focus on the potential bias that local moves could introduce in the context of free energy calculations.

We first look at staged AHFE calculations alternating global and local moves, to examine how much, if any, bias is introduced with the addition of local moves.

We followed up by evaluating local MD in the context of sequential Monte Carlo to determine bias as well as some measure of efficiency. 

\subsubsection{Mixed local and global sampling within staged calculations}
\label{section:staged_ahfe}
A standard~\cite{Matos2017-dl} approach to predict the AHFE of a compound is to perform staged equilibrium simulations, where each window is sampled using global MD.
We compare this default approach with an alternating sequence of local and global MD moves, assessing whether alternating between global and local moves provide unbiased AHFE predictions.

We selected a subset of FreeSolv~\cite{Matos2017-dl,Mobley2014-aw} (\url{https://github.com/MobleyLab/FreeSolv/releases/tag/v0.52}) to benchmark AHFE calculations using local moves for AHFE calculations, refer to Appendix \ref{appendix:freesolv_selection} for details.

\begin{figure}
    \centering
    \includegraphics[width=0.45\columnwidth]{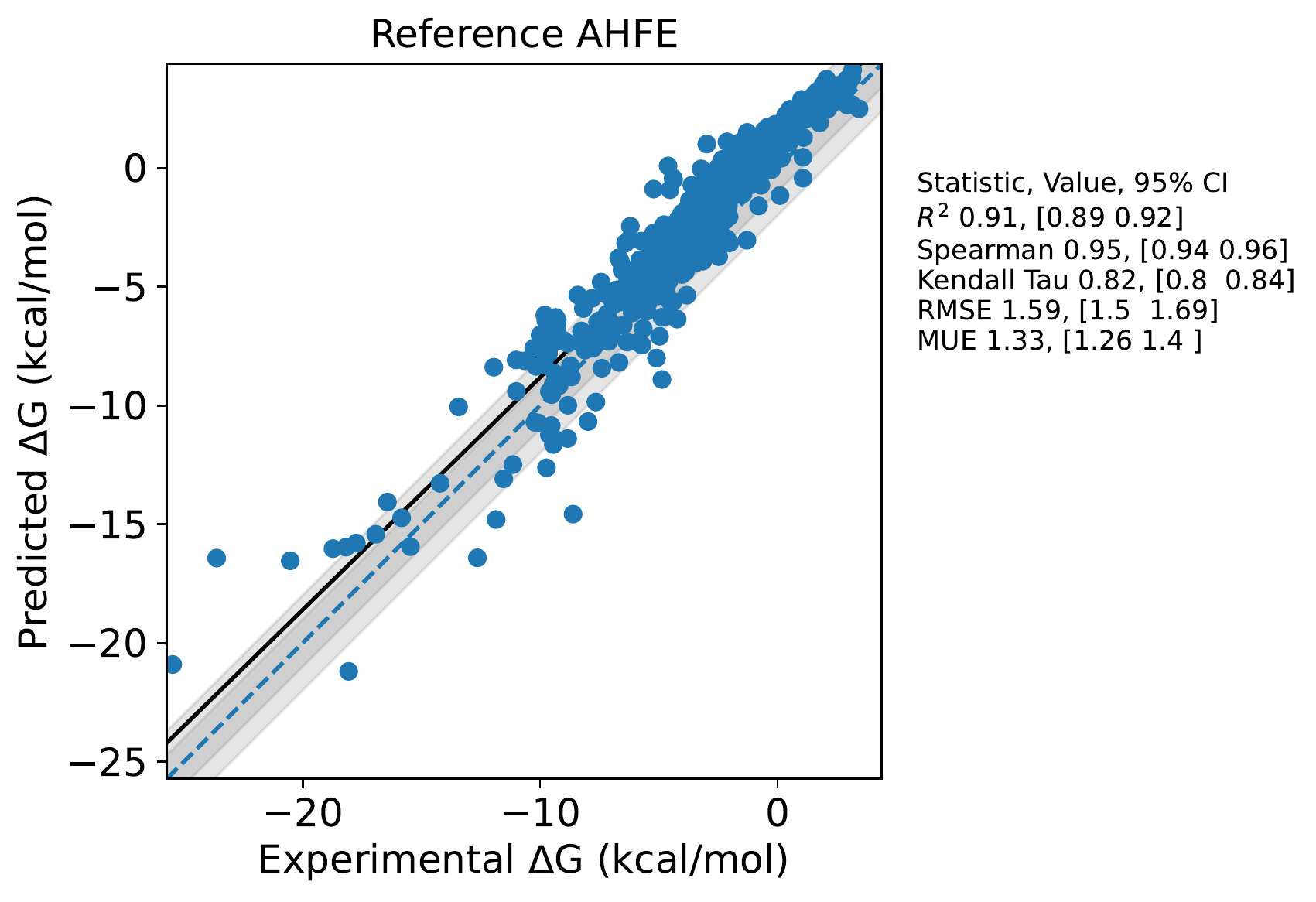}
    \includegraphics[width=0.45\columnwidth]{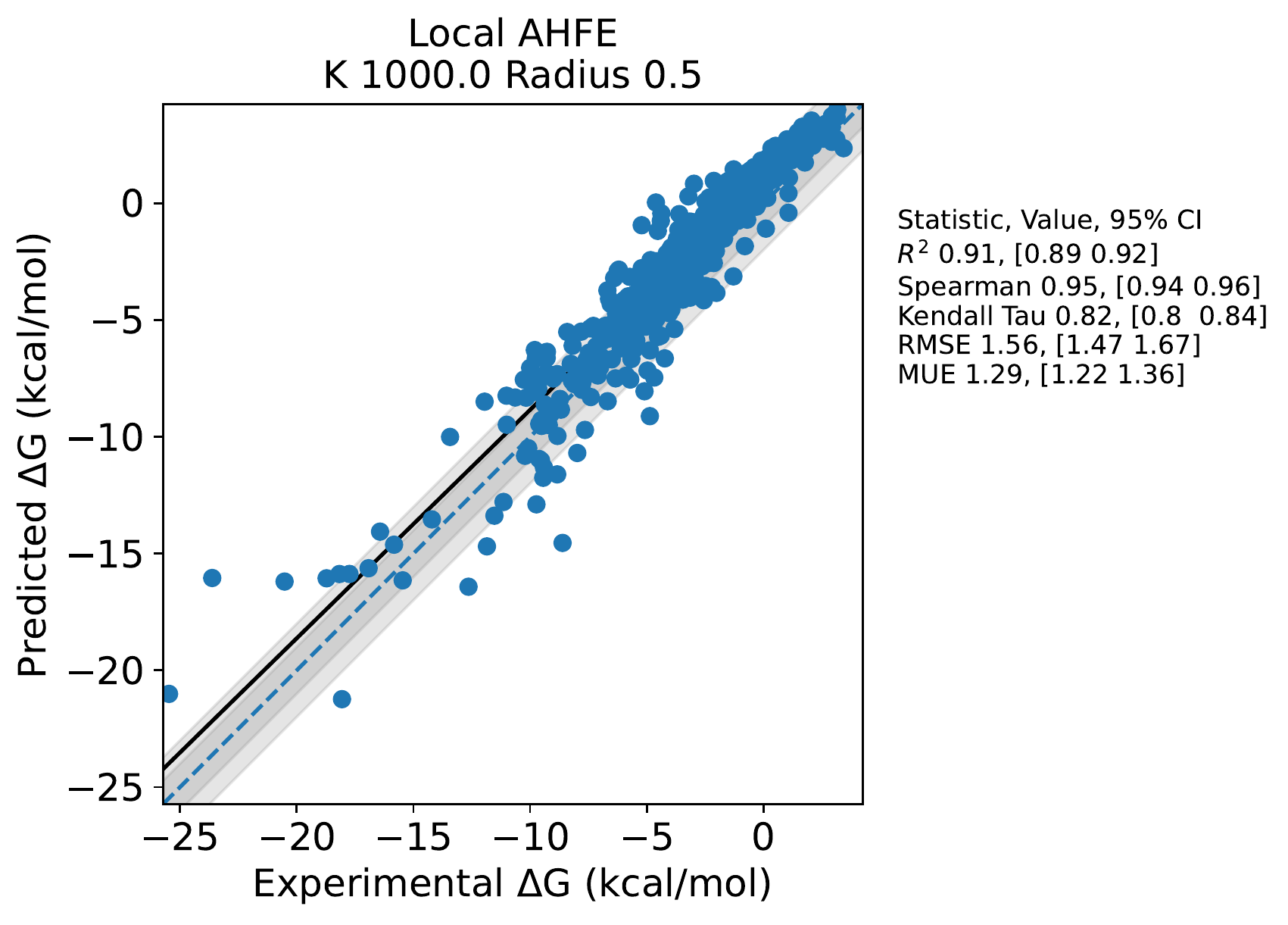}
    \includegraphics[width=0.45\columnwidth]{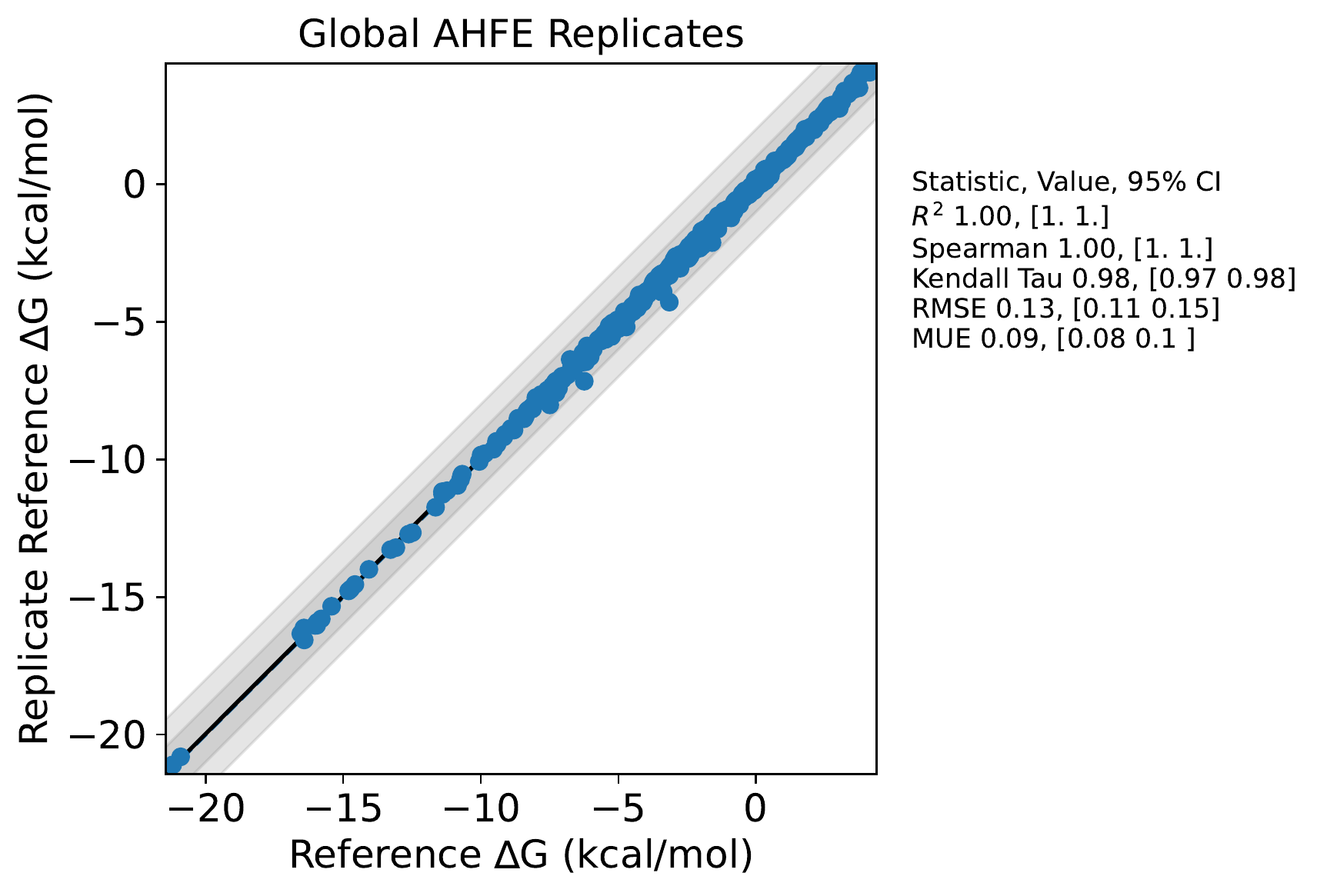}
    \includegraphics[width=0.45\columnwidth]{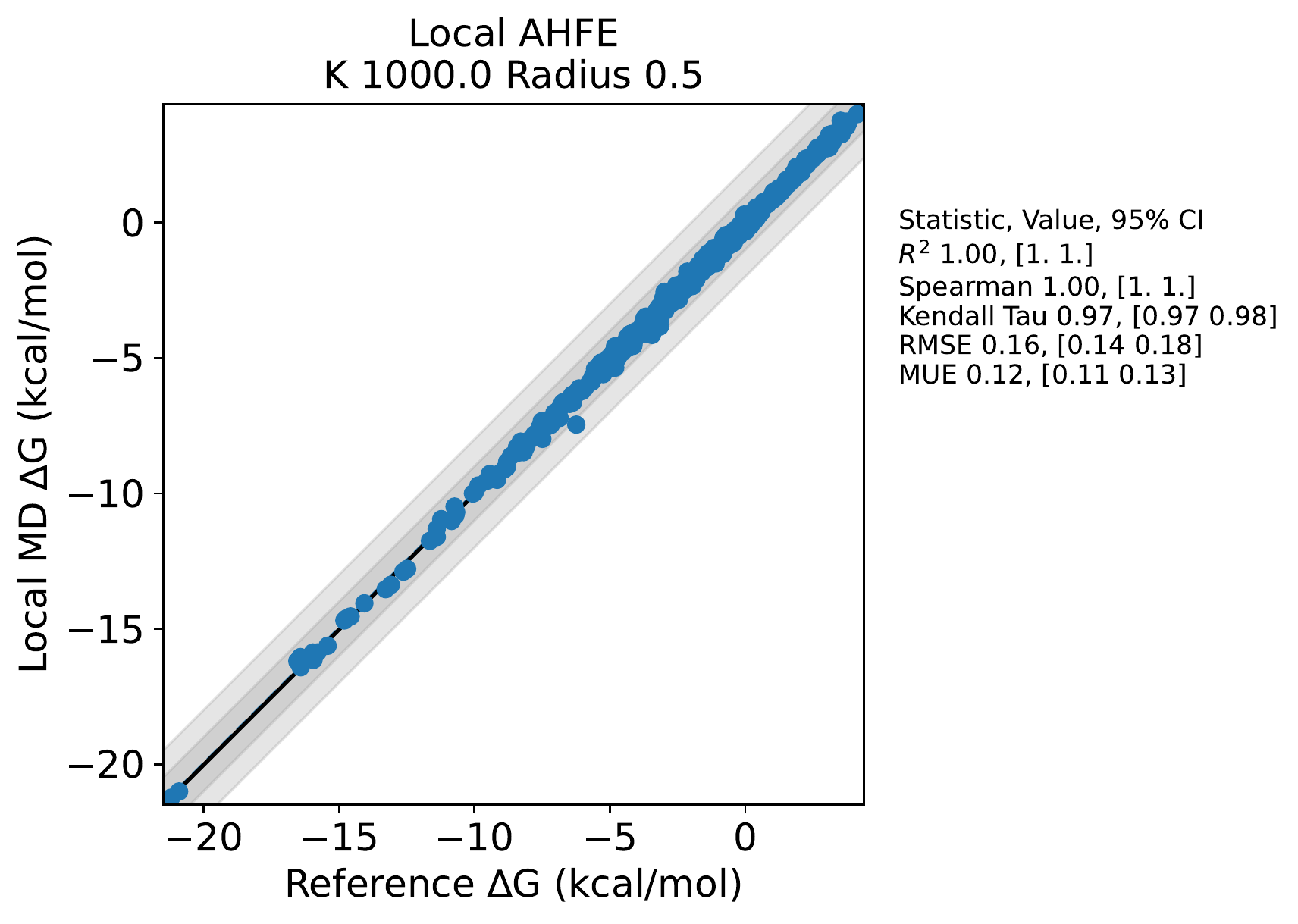}
    \caption{\textbf{AHFE validation on FreeSolv}
    Top left panel: Performance of the predicted AHFE using the reference protocol that uses the standard global MD approach.
    Top right panel: Performance of the AHFE predictions alternating global and local moves, which performs within the precision of the reference protocol.
    Bottom left panel: Compares predictions using global moves only between two replicates.
    Bottom right panel: Predictions using global moves vs. predictions using a mix of global and local moves.
    The outlier in the bottom right panel which shows greater than 1 kcal/mol difference has sampling issues, refer to Appendix \ref{appendix:staged_ahfe_slow_sampling} for more details. 
    }\label{fig:ahfe_comparison}
\end{figure}

\begin{figure}
    \centering
    \includegraphics[width=0.45\columnwidth]{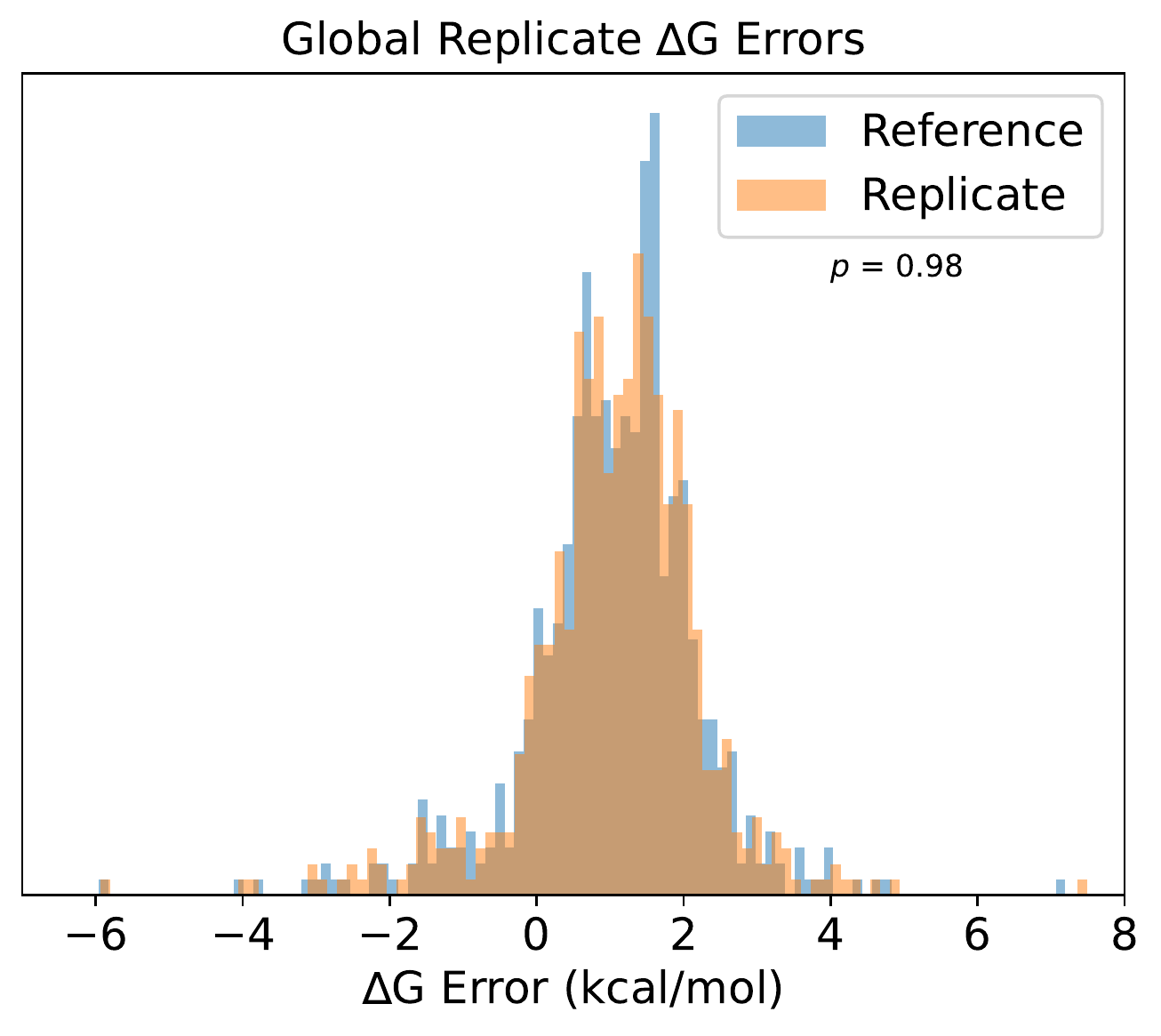}
    \includegraphics[width=0.45\columnwidth]{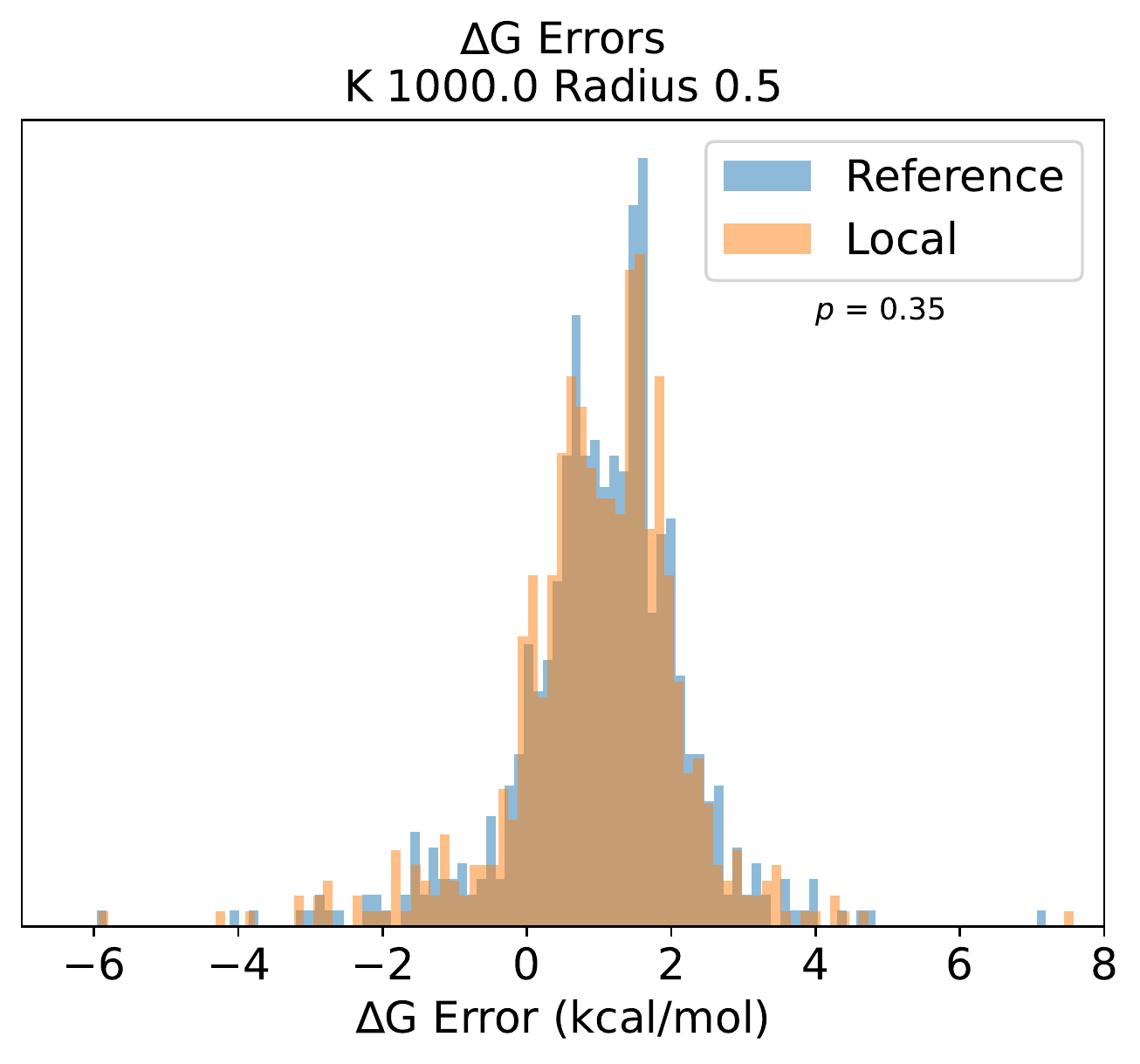}
    \caption{\textbf{AHFE residuals of the reference and local protocols.}
    First Panel: The residuals of the reference protocol compared to a replicate.
    Second Panel: The residuals of the reference run and the local run of AHFE, which shows a similar distribution.
    Using the two-sided t-test both result in a $p$-value that indicates that the differences in the distributions are not statistically significant.
    }\label{fig:ahfe_residuals}
\end{figure}

Using an identical system setup for the AHFE simulations alternating global and local and global MD using parameters that follow the rule of thumb described in Section \ref{subsection:local_with_unadjusted} ($k$ of 1000.0 kJ/mol / nm$^4$, radius of 0.5 nm, 500 local steps and 100 global steps) results were within the precision of the reference AHFE protocol, as summarized in Figures \ref{fig:ahfe_comparison} and \ref{fig:ahfe_residuals}.
For complete simulation details, refer to Appendix \ref{appendix:staged_ahfe_details}.
The results suggest  no significant bias is introduced by alternating global and local moves in the context of explicit solvent staged calculations.

\begin{figure}
    \centering
    \includegraphics[width=0.9\columnwidth]{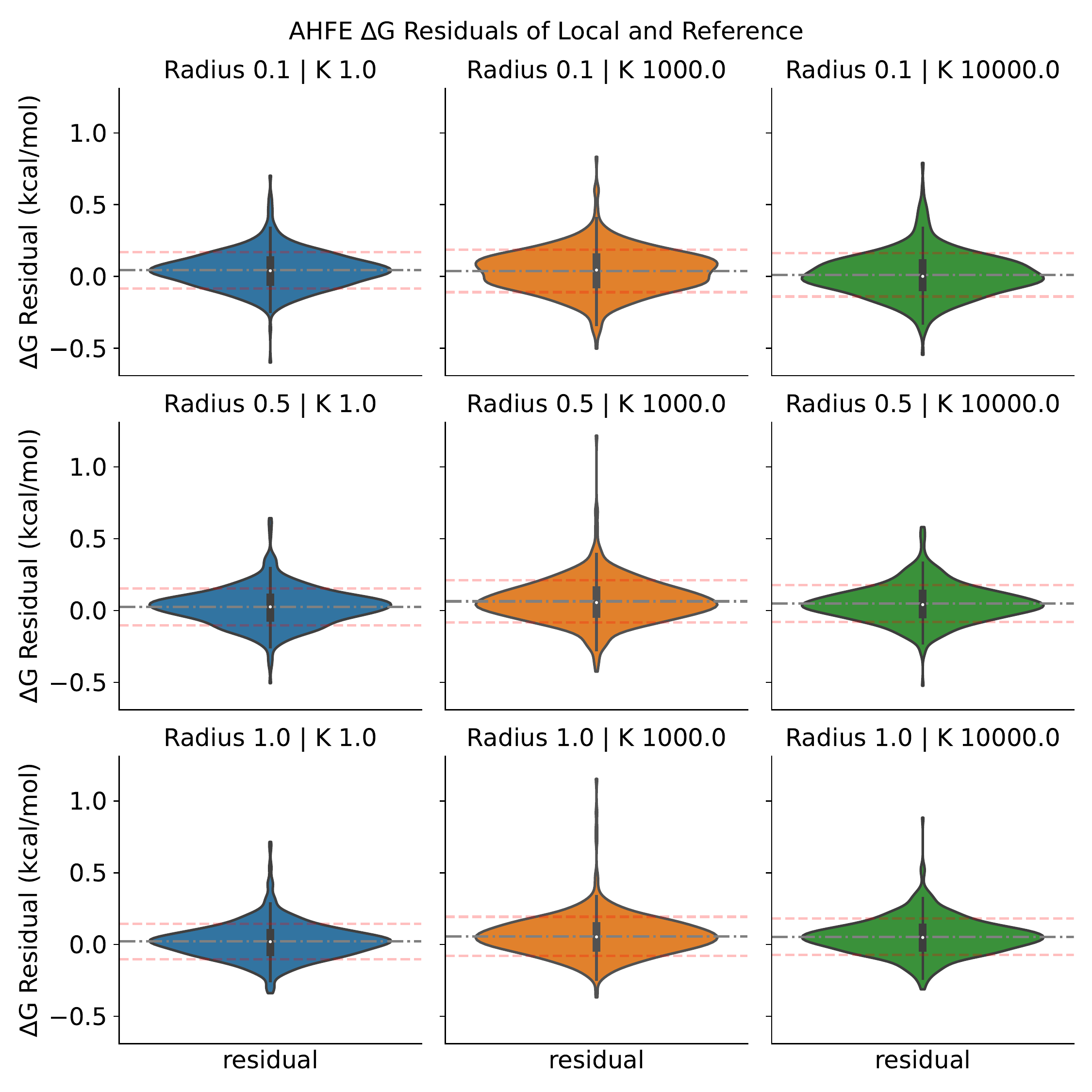}
    \caption{\textbf{$\Delta G$ residuals for staged AHFE calculations of the local protocol compared to the reference protocol.}
    The residuals of the predicted reference $\Delta G$ - local $\Delta G$ are distributed normally and are within the precision of the reference method.
    The residuals display the error distribution in kcal/mol for each combination of radius and $k$ tested, demonstrating very similar performance across parameters tested.
    Radius values are in units of nanometer and $k$ are in kJ/mol/nm$^4$. 
    Figure \ref{fig:appendix_local_to_ref_all} in the Appendix has plots comparing the reference $\Delta G$ values to the local protocol for all parameters displayed here.
    The dash dot lines and dashed lines in the violin plots are the mean and standard deviation of the residuals respectively.
    }\label{fig:local_ahfe_errors}
\end{figure}

AHFE predictions were run using 9 combinations of restraint parameters: three values of restraint strength $k$ (1.0, 1000.0, 10000.0) and three radii (0.1nm, 0.5nm, 1.0nm) in an attempt to observe how these parameters impacted free energy predictions, refer to  Appendix \ref{appendix:staged_ahfe_details} for more details.
The difference in the predicted $\Delta G$ of the protocol which mixes local and global steps to the reference protocol across the parameter space shows little evidence of sensitivity to radius or restraint strength, as summarized in Figure  \ref{fig:local_ahfe_errors}.
This insensitivity suggests little bias is introduced by the radii and restraint strength in an explicit solvent environment.

\subsubsection{Local sampling within sequential Monte Carlo}
\label{section:adaptive_smc}

In Section \ref{section:staged_ahfe}, we controlled the number of steps performed, and confirmed that alternating between unadjusted global and local MD moves does not introduce significant bias into staged hydration free energy predictions.
However, it is difficult to assess the impact of including local moves on overall sampling productivity in this setting.

When controlling for the possibly lower productivity of local MD moves, do application-level efficiency gains still remain?
We attempted to assess this within a framework that adjusts the number of sampling steps based on a measure of their productiveness.
One such framework is adaptive sequential Monte Carlo (SMC)~\cite{Suruzhon2022-gt, Everitt2020-do, Dai2022-ke}, where the next $\lambda$ window can be selected adaptively based on the current samples and a heuristic that is sensitive to both the problem’s difficulty and the chosen sampling method’s effectiveness.
More details are in Appendix \ref{appendix:smc_details}, but note that this adaptive selection:
\begin{enumerate}
    \item Depends on a parameter that governs the precision (chosen arbitrarily as $\beta = 0.95$).
    This controls how much the effective sample size (ESS) is allowed to decrease between one window and the next $\text{ESS}_{t+1} = \beta \text{ESS}_t$.
    \item Introduces additional overhead to each window’s computation, further limiting the achievable speed-up.
    (In addition to MD, we must evaluate energies on all stored snapshots several times at trial values of $\lambda$ while searching for the next $\lambda_t$.)
\end{enumerate}

\begin{figure}
    \centering
    \includegraphics[width=1.0 \columnwidth]{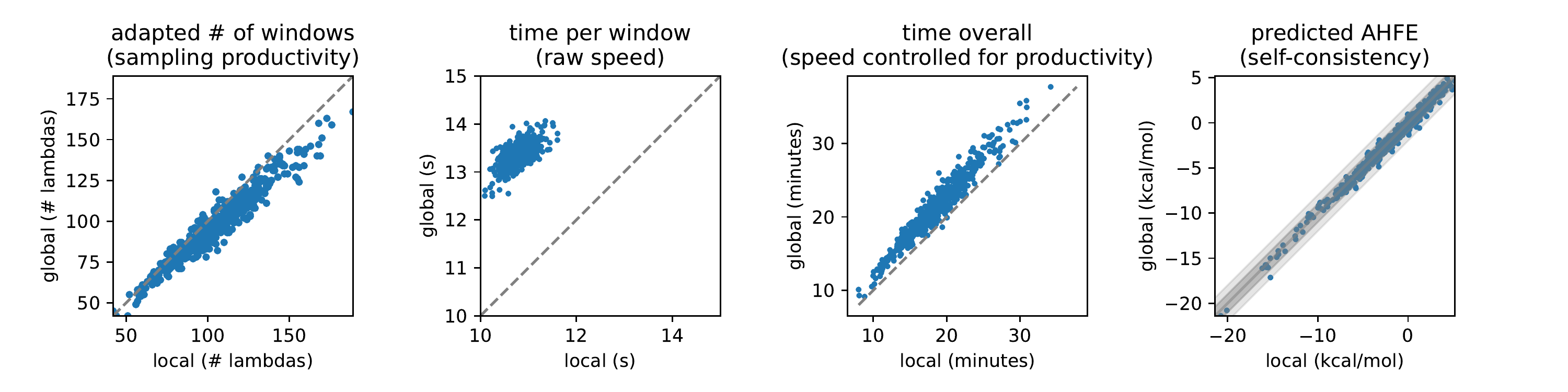}
    \caption{
    \textbf{Local propagation modestly accelerates an application where differences in sampling productivity are controlled for.}
    Each dot represents a single molecule in FreeSolv.
    First panel: Using local propagation (radius = 1 nm) has lower productivity per step, slightly increasing the number of $\lambda$ windows selected by adaptive SMC.
    Second panel: However, the time per window is reduced by a larger factor.
    Third panel: The net effect is a modest application-level speed-up.
    Fourth panel: SMC with local propagation produces similar predictions to SMC with global MD propagation.
    }
    \label{fig:smc_performance}
\end{figure}

Applying adaptive SMC to AHFE prediction, holding all settings fixed except the choice of using global vs. local MD for propagation, we can measure the competing factors of decreased sampler productivity per step, vs. increased computational speed up per step.
In this application, global and local moves are not alternated within the same simulation as in Section \ref{subsection:intro_local_md}, but instead a simulation is configured to use only local MD moves (250 steps per window), or a simulation is configured to use only global MD moves (250 steps per window).
Adaptation increases the overall number of $\lambda$ windows for local MD compared to global MD (about 10\% on average) in this case, but this is more than compensated for by the increased speed per window when using local MD moves (about 30\% on average), resulting in an application-level speed up of about 13\% for local moves of radius = 1 nm, as summarized in the first three panels of Figure \ref{fig:smc_performance}.
In the context of expected scaling and preliminary practical scaling (discussed later in Section \ref{section:practical_speed_up}), 
a larger speed-up may be expected for applications where the simulation box is larger, and as further performance optimizations are applied.

We also confirm that SMC using local MD propagation results in comparable predictions on FreeSolv (97\% within 1 kcal/mol of each other), shown in the final panel of Figure \ref{fig:smc_performance}.
This suggests that the uncontrolled bias associated with using local MD in this setting is small.
Note that the sampling biases in the context of SMC may be different from the sampling bias that may arise from alternating unadjusted global and local MD moves (which was investigated in Section \ref{subsection:local_with_unadjusted}).


\subsection{Relative binding free energy}
\label{section:staged_rbfe}

In the previous sections (\ref{section:staged_ahfe} and \ref{section:adaptive_smc}) we considered applications of local MD moves in the context of an explicit solvent environment.
This validates that local moves can be a practical addition -- introducing no discernible bias into free energy estimates, and realizing a practical speed-up.

Next, we consider applications in the context of binding.
Similar to the previous staged calculation in Section \ref{section:staged_ahfe}, we wish to validate that including local moves does not introduce much bias into free energy estimates, and to assess the practical speed-up.
We also wish to assess qualitatively if the radial selection moves developed in Section \ref{section:radial_selection} are appropriate for ligand sampling in the presence of a protein.

We use JACS Tyk2 benchmark set~\cite{Wang2015-dx} for our evaluation of staged RBFE calculations alternating global and local moves.
Tyk2 was chosen since it is a standard benchmark and had shown to have small variance under the reference protocol that we compared against alternating between global and local MD.

\subsubsection{Alternating global and local moves}
\label{section:rbfe_global_local}
In Section \ref{section:staged_ahfe}, we evaluated alternating global and local moves to compute free energies in an explicit solvent environment, in this section we extend our analysis to RBFE prediction. 

To evaluate the efficiency and bias of alternating global and local moves in RBFE predictions, we ran Tyk2 under three different conditions:
\begin{enumerate}
    \item The reference protocol of 400 global steps per frame,
    \item 100 global steps followed by 300 local steps per frame, and
    \item 100 global steps per frame.
\end{enumerate}
Each condition collected a total of 2000 frames, with only the number of steps and the types of steps between frames being the difference.
At the beginning of each start of the 300 steps in the second condition, a reference particle was selected randomly from the ligand particles in the edge, which was used to defined the free particles for the local moves.
These three conditions allow the exploration of the bias introduced by alternating global and local steps, as well as how much efficiency is gained with the additional local steps.
Refer to Appendix \ref{appendix:rbfe} for additional details on simulation setup.

\begin{figure}
    \centering
    \includegraphics[width=0.3\columnwidth]{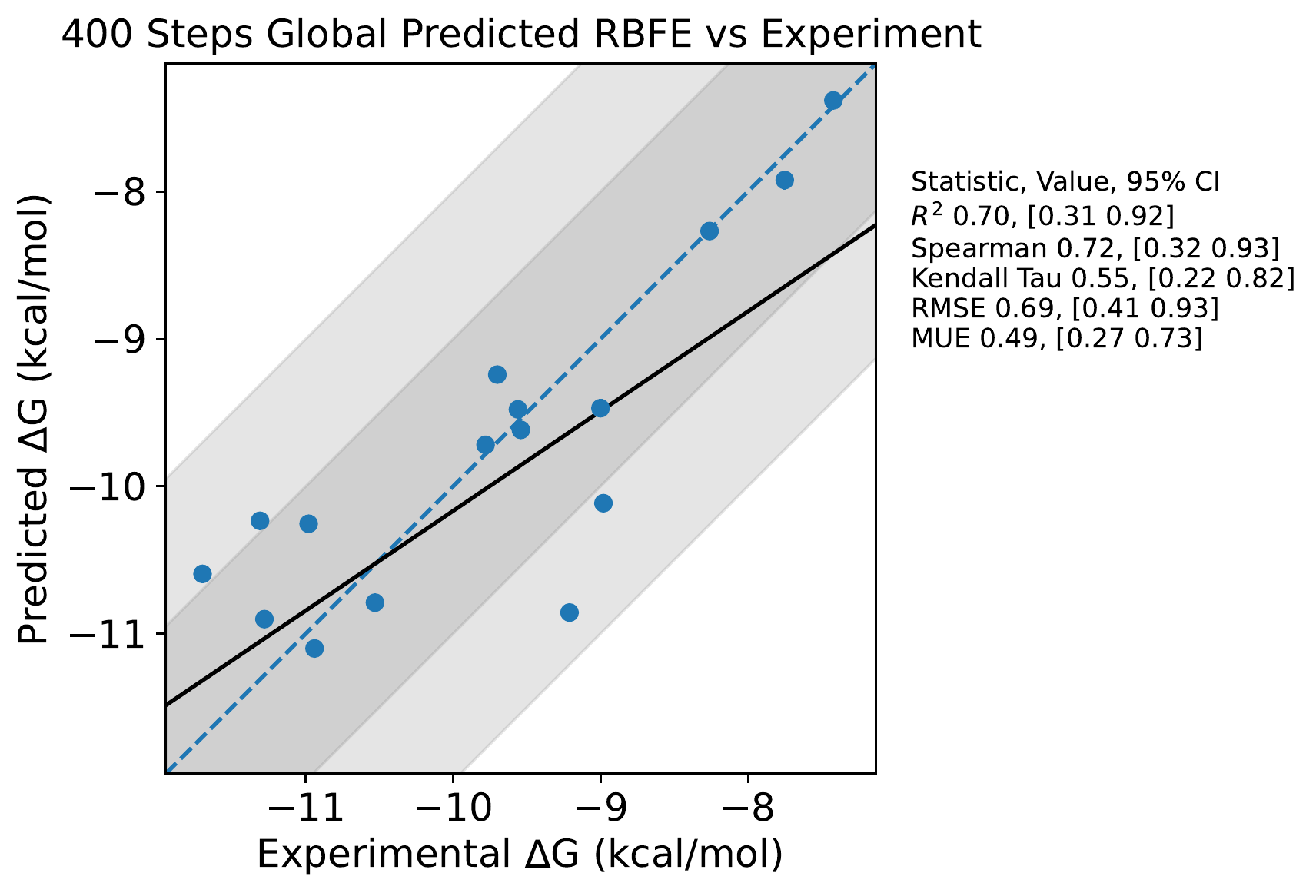}
    \includegraphics[width=0.3\columnwidth]{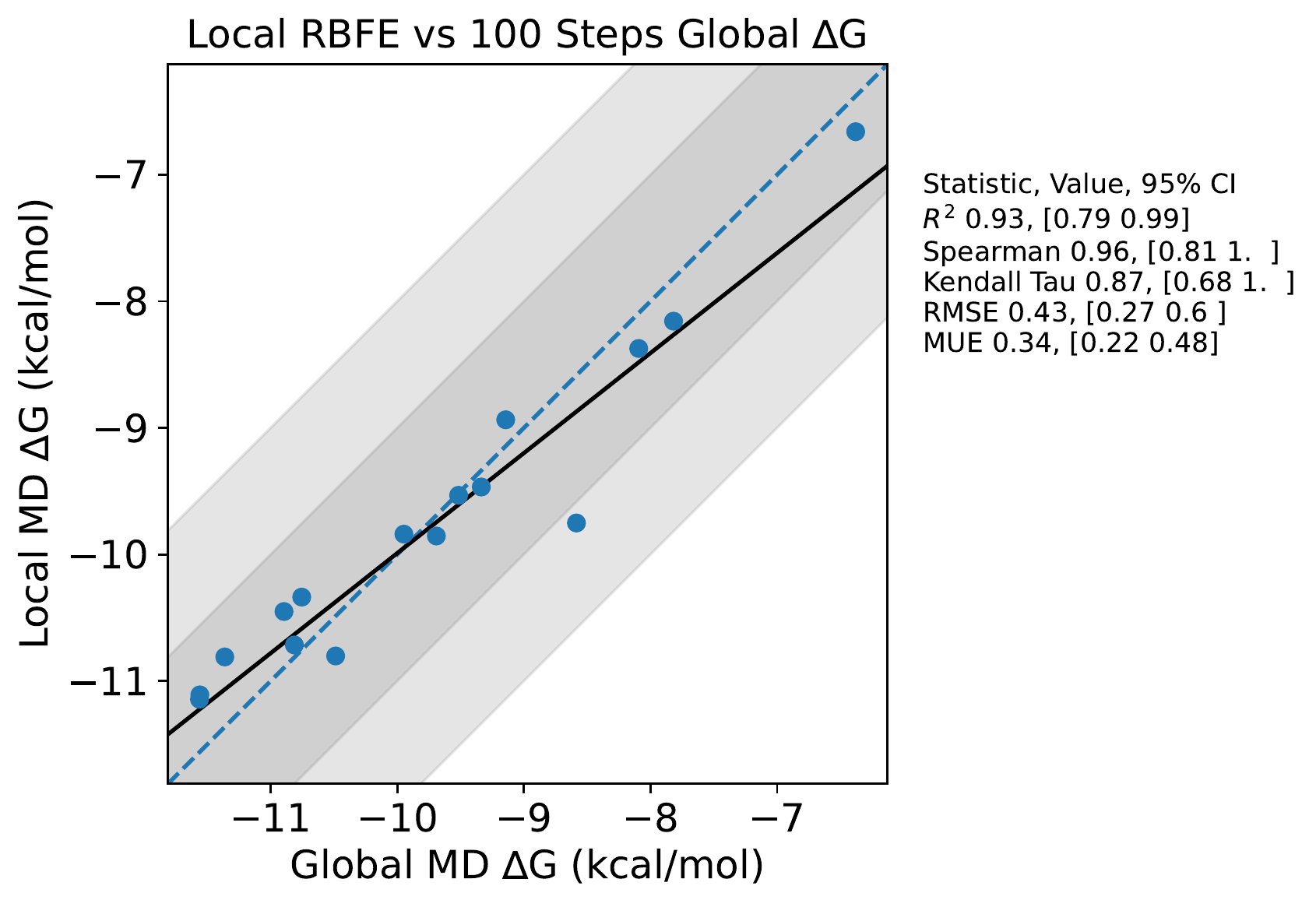}
    \includegraphics[width=0.3\columnwidth]{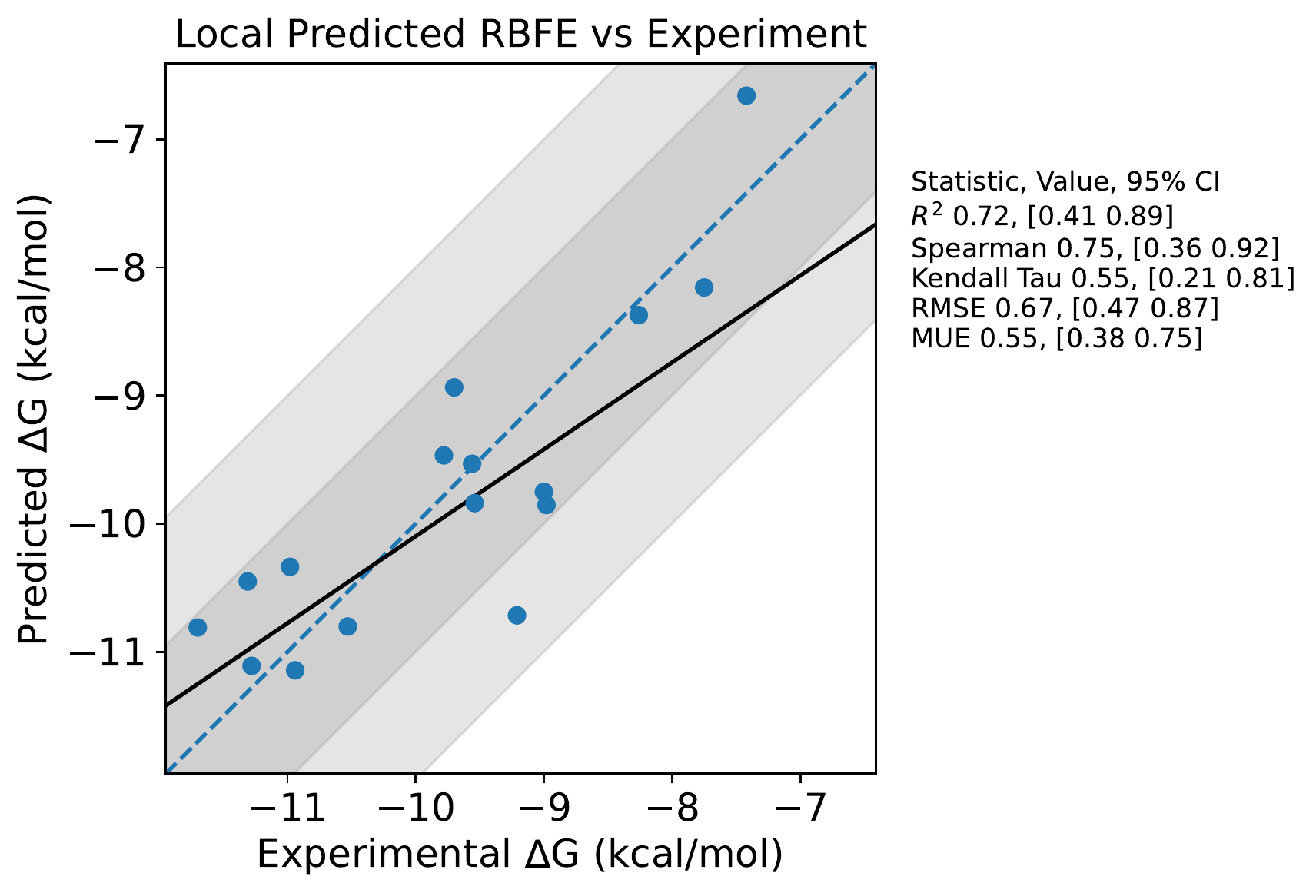}
    
    \caption{\textbf{RBFE predictions of Tyk2 with and without local moves.}
    First panel: Accuracy of predicted $\Delta G$s of RBFE calculations of the reference protocol of 400 steps per frame.
    Second panel: Accuracy of predicted $\Delta G$s of RBFE calculations with only 100 global steps per frame.
    Third panel: Accuracy of predicted $\Delta G$s of RBFE calculations with 100 global steps followed by 300 local steps per frame.
    }\label{fig:rbfe_accuracy}
\end{figure}

\begin{figure}
    \centering
    \includegraphics[width=0.3\columnwidth]{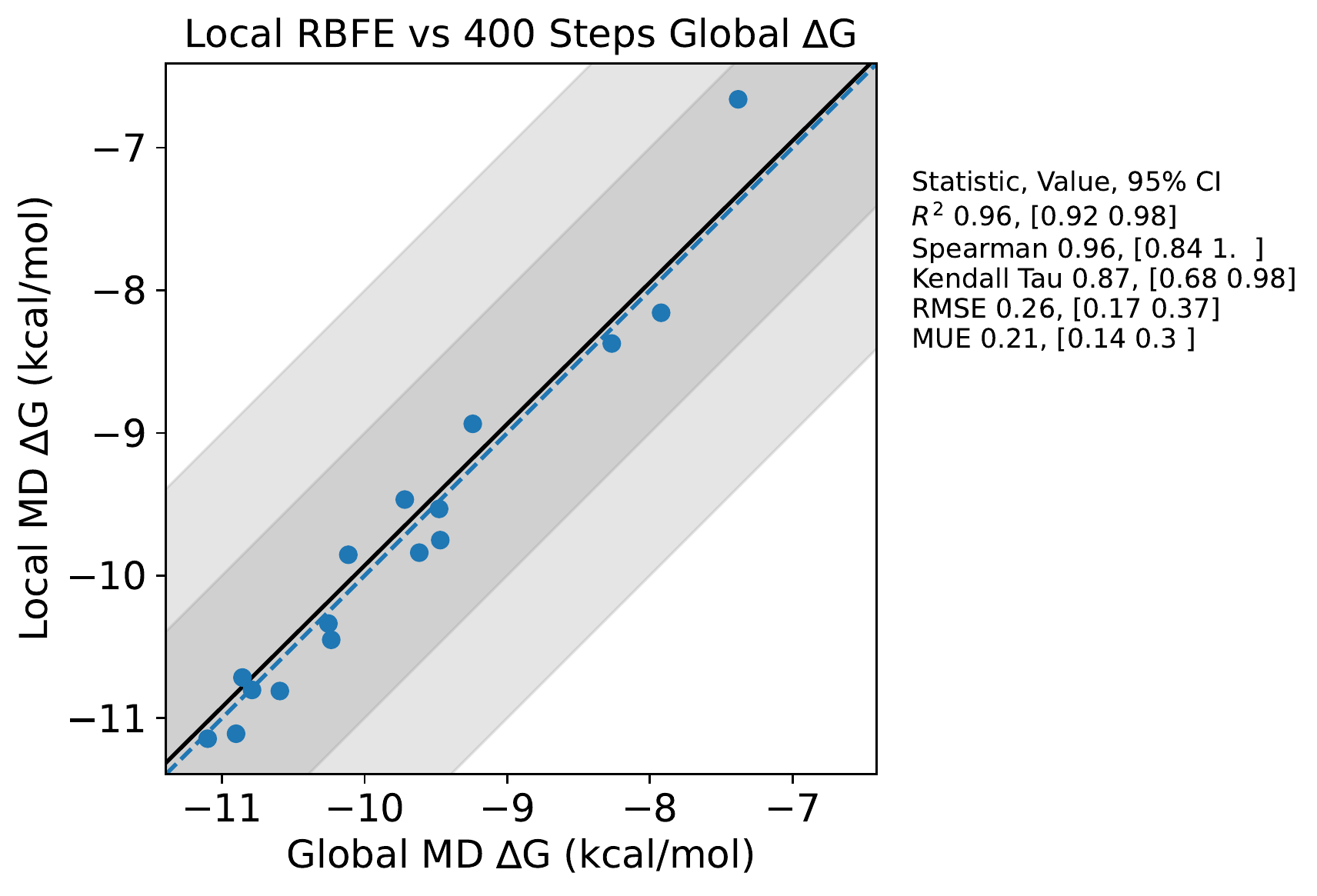}
    \includegraphics[width=0.3\columnwidth]{figures/rbfe/rbfe_local_vs_ref_100.pdf}
    \includegraphics[width=0.3\columnwidth]{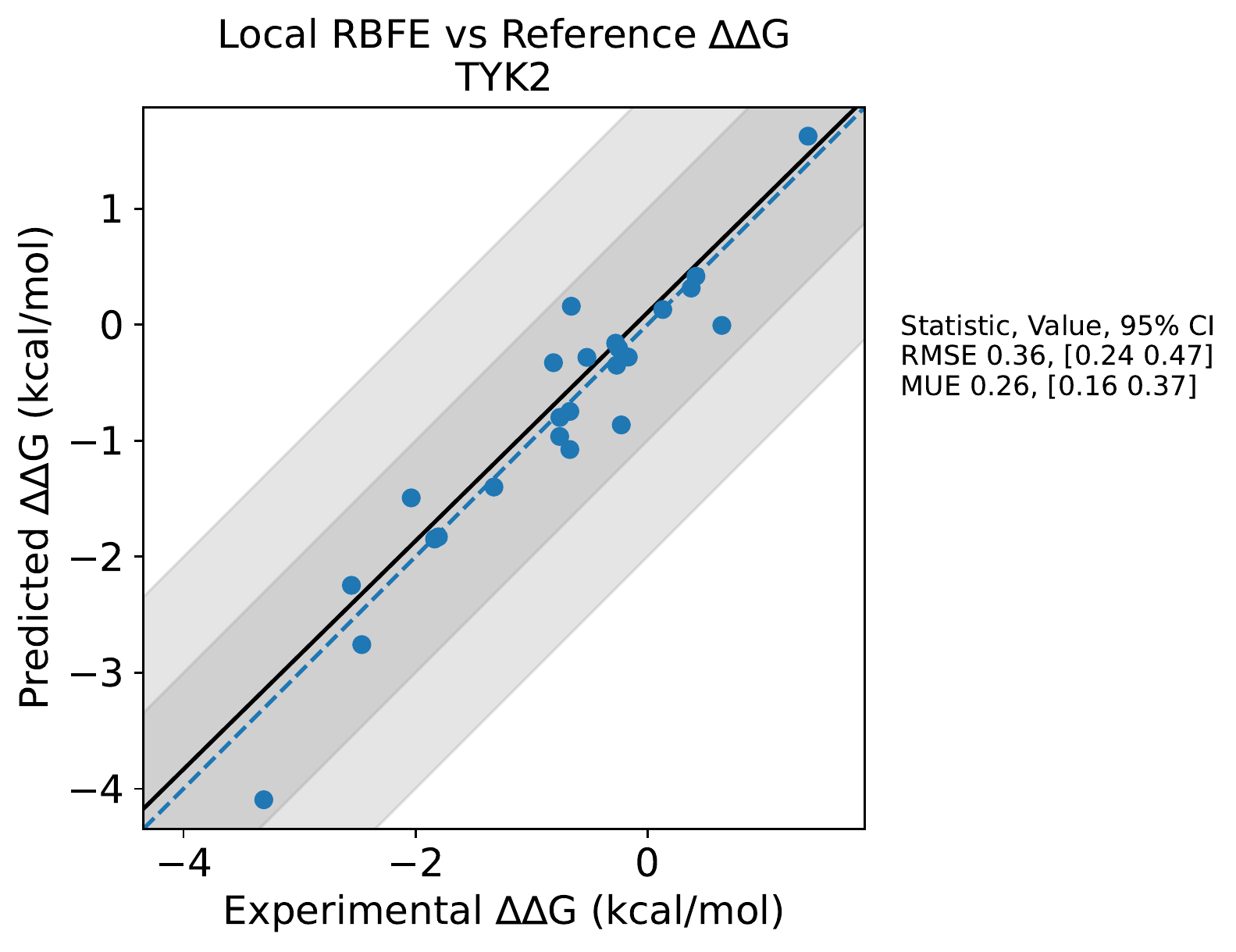}
    \caption{\textbf{Comparison of RBFE protocols with and without local steps.}
    First panel: Comparing the $\Delta G$s of RBFE calculations with 100 global steps followed by 300 local steps to that of the reference protocol with 400 global steps.
    Second panel: Comparing the $\Delta G$s of RBFE calculations with 100 global steps followed by 300 local steps to that of only running 100 global steps per frame.
    Third panel: Comparing the $\Delta\Delta G$s of RBFE calculations with 100 global steps followed by 300 local steps to that of the reference protocol.}
    \label{fig:rbfe_comparisons}

\end{figure}

The accuracy of the three protocols for predicting RBFE on Tyk2 produced similar results, as summarized in Figure \ref{fig:rbfe_accuracy}.
This is to be expected, as local moves should not impact the accuracy of the measure, only impact the potential bias in the measurement.
When comparing the different protocols against one another, no measurable bias was discernible when compared to the protocols with only global moves, as seen in Figure \ref{fig:rbfe_comparisons}.
This bolsters our confidence that local moves introduce little bias when mixed appropriately with global steps.

In the next section, we evaluate the impact of local MD parameters (radius, $k$) on free energy calculations on a single edge.

\subsubsection{Ablation study and radius sweep}
\label{section:rbfe_only_local}

The previous section (\ref{section:rbfe_global_local}) suggests that introducing local moves (with a specific radius) introduces little bias across the Tyk2 benchmark set.
Two natural follow-ups are: How small can the radius be made in the context of binding, and how sensitive is this result to the inclusion of a baseline amount of global MD sampling?

To address this, we also run a staged calculation where each window is initialized with a structure equilibrated by global MD, but production sampling relies entirely on local MD.
We stress that the simulation setup in this section is not expected to converge to the correct answer in general, since local MD moves are not ergodic.
To converge to the correct answer in the limit of infinitely long sampling, some ergodic moves need to be included, or local moves can be applied in a context that does not assume ergodic sampling (such as nonequilibrium switching or SMC, rather than staged calculations).

\begin{figure}
    \centering
    \includegraphics[width=0.8\columnwidth]{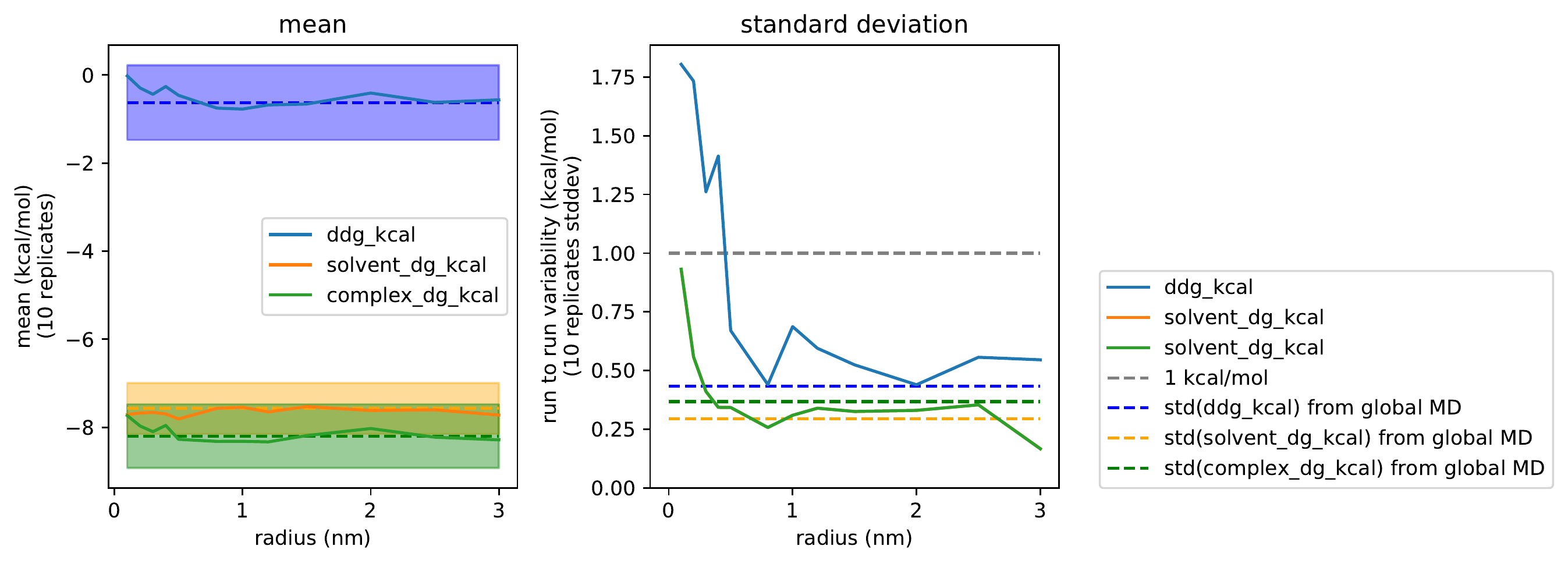}
    \caption{\textbf{Bias and precision of staged RBFE calculations on a single edge, with fixed-radius local moves.}
    First panel: Mean predictions of solvent $\Delta G$, complex $\Delta G$, and binding $\Delta \Delta G$ as a function of local move radius.
    Second panel: Standard deviation of 10 replicates.
    }
    \label{fig:tyk2_replicates}
\end{figure}

We selected a single edge from the JACS Tyk2 benchmark set, arbitrarily selected by picking the edge (``jmc23'', ``ejm55'') with the highest absolute value of experimental $\Delta\Delta G$.
We computed 10x replicates of this edge at a grid of local radii, with varying random seeds, and assessed bias and run-to-run variability in Figure \ref{fig:tyk2_replicates}.
We also computed 10x replicates using conventional MD for the same number of steps.
These results suggest that a larger effective radius is required for complex than for solvent leg, and the estimation bias even in this case is small.

However, this is a single transformation in the context of a single target -- requirements are likely to vary by target and by transformation.
For example, if global protein reorganization contributions are significantly different between the two ligands, local sampling alone is unlikely to capture this difference efficiently.


\section{Discussion}
\label{section:discussion}

\subsection{Quantifying theoretical and practical speed-ups}
\label{section:practical_speed_up}
The method described above provides a way to extract small subproblems from a larger sampling problem.
This can be expected to reduce the number of interactions that \emph{need} to be computed per sampler step, but this does not automatically translate into a corresponding speed-up in practice, even in the context of a simulation involving short-range pair potentials.

Roughly, there are two ways this could yield a speed-up in practice: making each force calculation cheaper, or reducing the number of force calculations.

\begin{figure}
    \centering
    \includegraphics[width=0.9\columnwidth]{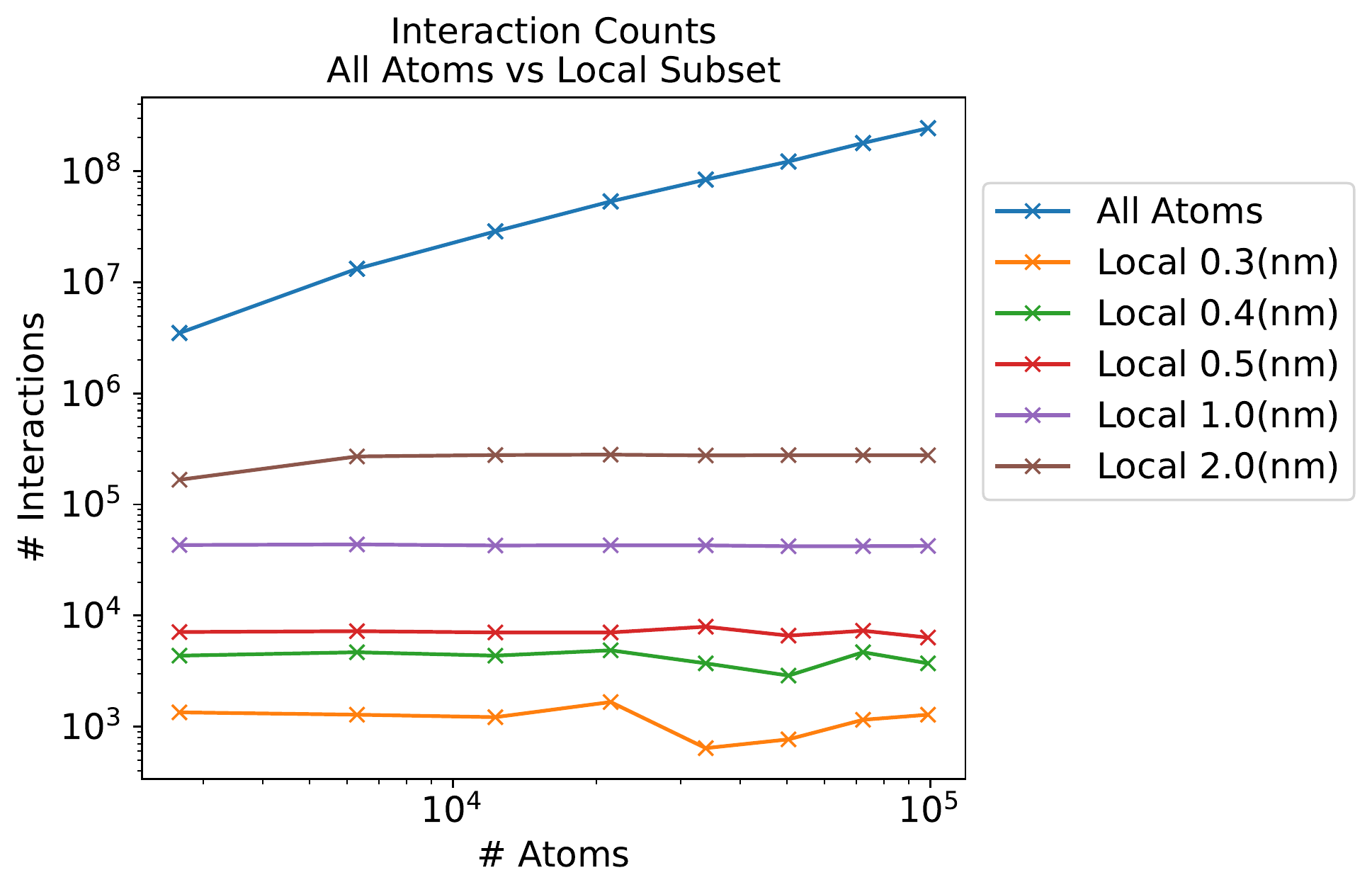}
    \caption{\textbf{The number of nonbonded interactions of global MD vs local MD regions.}\\
    Compares the number of nonbonded interactions of running global MD computed on the entire system using a 1.2nm cutoff vs the number of interactions based on the region simulated with local MD within a certain radius.
    For the local regions, the interactions are between the particles within the region and the interaction of the internal region with those within 1.2nm outside the region.
    Local MD regions produce near constant numbers of interactions, which presents the best possible speed up possible from a local MD implementation.
    Shown local region is based on a benzene in water boxes of various sizes. 
    Note that the number of interactions is dependent on the density of the system.
    }
    \label{fig:tile_speedup}
\end{figure}

\begin{figure}
    \centering
    \includegraphics[width=0.9\columnwidth]{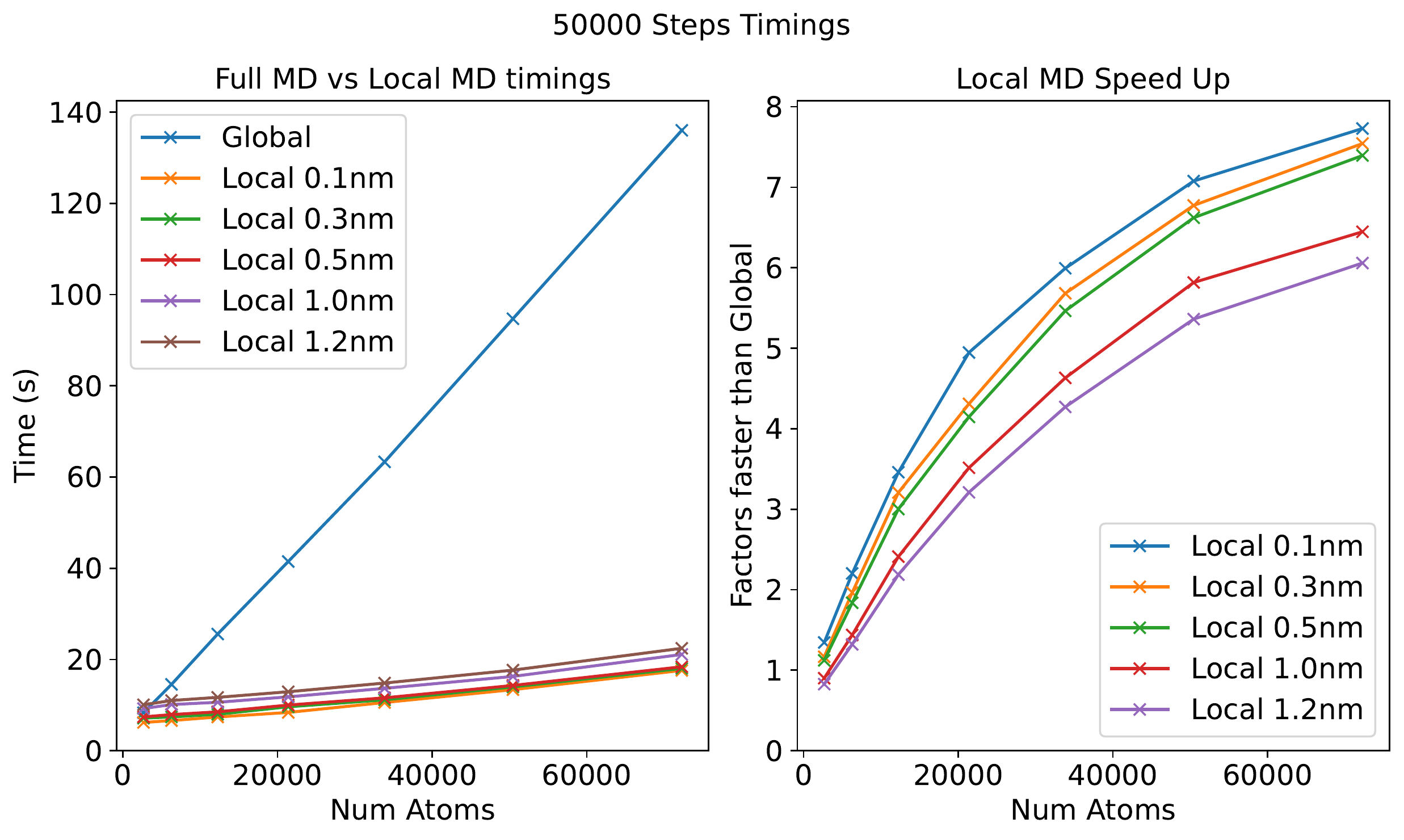}
    \caption{\textbf{Realized performance gain increases as a function of system size.}\\
    First panel: wall time (in seconds) to simulate 50,000 steps of global MD, or local MD with selection radius of 0.1, 0.3, 0.5, 1.0, or 1.2 nanometers, restraint strength $k$ = 1000 kJ/mol/nm$^4$, for boxes of TIP3P water systems containing between 10,000 and 90,000 atoms.
    Time per step of global MD increases linearly with system size, local MD increases linearly as well but with a much smaller constant with respect to system size.
    Second panel: Factor speed-up (global MD time / local MD time).
    Simulation details are described in Appendix \ref{appendix:benchmark_simulation_details}
    }
    \label{fig:speed_benchmark}
\end{figure}

The maximum expected cost savings per force calculation depends on the size of the system, the size of the subsystem, and what interactions are modeled.
In the idealized case of a homogeneous system containing $N$ atoms interacting withing a short-range atomic cutoff $r_c$, the total number of interactions is approximately $O(N r_c^3)$.
The idealized expected speed-up would be linear: $O(\frac{N r_c^3}{N_\text{selected} r_c^3}) = O(\frac{N}{N_\text{selected}})$.
Given a fixed set of local parameters, the number of interactions will be constant for all system size, as shown in Figure \ref{fig:tile_speedup}.
This will provide a speed-up relative to the system size, with smaller systems expected to have less benefit from local sampling.
The currently realized speed-up is shown in Figure \ref{fig:speed_benchmark}.

Note that achieving the optimal linear speed-up in practice is highly dependent on implementation details such as the data structure of neighborlist, the voxelization procedures, and various overhead associated with kernel execution. In addition, fully utilizing the bandwidth of modern hardware may require additional amortization of overhead by means of batching.
However, in most MD engines, the dominant bottleneck is evaluation of the nonbonded energy, typically computed as a sum of short-range and long-range interactions. 
For short-range interactions, this will require modifying the neighborlist algorithm to work efficiently on the selected atoms and skipping evaluations between de-selected particles. However, for long-range interactions, assuming a periodic system evaluated under particle mesh Ewald (PME)~\cite{Essmann1995-dj}, a major limitation is that a local perturbation in direct space generally results in a global perturbation of the reciprocal space structure factors:

\begin{equation}
\begin{split}
S(\mathbf{k}) &= \sum_j^N q_j \exp(i \mathbf{k} \cdot \mathbf{x_j}) \\
 &= \sum_j^N q_j \left( \cos(\mathbf{k} \cdot \mathbf{x_j}) + i \sin(\mathbf{k} \cdot \mathbf{x_j}) \right) \\
\end{split}
\end{equation}

While it may be possible to more efficiently evaluate a particular structure factor, the number of structure factors is still dependent on the PME grid size. The dimensions of the PME grid is typically proportional to the volume, which is indirectly proportional to the total number of particles in the system. However, the method may still be applicable in the context of multiple time-step integrators where the long-range terms are evaluated less frequently~\cite{Zhou2001-cj}. 



Additionally, there may be settings where a practical speed-up can be realized even without making the force computations themselves more efficient.
For example, restricting attention from the entire simulation to a local selection may reduce the number of steps required to compute an estimate to a given precision.
A conservative example is considered in Appendix \ref{appendix:local_mcmc}, where HMC moves applied to local selections are found to tolerate larger timesteps than HMC moves applied to the entire system.
This avenue may provide larger benefits in the context of sampling methods or estimators that scale poorly with the problem dimension.

\subsection{Limitations and future work}

\subsubsection{Selection of the focus region}
\label{section:better_selection}

We elaborated one sequence of choices that leads to a convenient implementation based on radial selection in Section \ref{section:radial_selection}, but this specific implementation may not be appropriate for all systems.

There is no fundamental reason the selected region has to be spherical, for example.
It could also be an irregular region such as ``anywhere near the surface of the protein'' (as long as the region can be expressed using a suitable restraint potential).


There may also be applications where rearranging particles selected in a spherical region near a fixed central particle is unproductive.
We stress that local moves as defined in Section \ref{section:local_resampling_recipe} are valid, but not necessarily productive or ergodic.
For example, the specialization in Section \ref{section:radial_selection} could be unproductive if the selected central index is on the ligand, if the selected sphere is small, and if productive moves require that the ligand as a whole translate relative to the binding pocket.
In such cases, the ``frozen reference'' particle could prevent translation of the ligand relative to the binding pocket.
This effect can be partly mitigated by
(1) reverting the choice to freeze the central particle (considered in Appendix \ref{section:freezing_central_particle}).
(2) periodically randomizing the selection of reference particle within the ligand,
(3) allowing to select binding pocket indices, or
(4) alternating local moves with global moves that can translate the ligand in the binding pocket.

In general if the local region and environment are tightly coupled, but local moves are implemented in such a way that they slow down sampling of environment fluctuations, then the proposed method can lead to slower sampling of fluctuations in the local region.
Revisiting the specializations in Section \ref{section:radial_selection} may lead to more productive local moves tailored to applications where this is a concern.

\subsubsection{Holonomic constraints}
In this work, we did not consider handling constraints.
Molecular simulations often introduce constraints to increase the maximum stable timestep, esp. by treating the waters as rigid bodies or by preventing vibration of certain bonds that would limit the maximum stable timestep.
In principle, we expect the proposed method should be compatible with such constraints, but there may be subtle implementation details to consider.
For example, if the distance $d_{ij}$ between particles $i$ and $j$ is constrained, but we only select one of the particles $i$ to be mobile, then the implementation of the constraint projection should not move the particle $j$.

\subsubsection{Varying granularity of selection}
This work did not consider grouping the problem variables by \emph{molecule} (or \emph{residue}) rather than by \emph{particle}.
This should be possible in principle, by partitioning the system into groups $i$, and having the boolean variable $z_i$ represent the selection of a group $i$, rather than a single particle $i$.

\subsection{Related work}
\label{section:related_work_discussion}

This work shares motivation with many other strains of work in molecular simulation, and draws on prior work in molecular simulation and computational statistics.
Some closely related work in molecular simulation includes ``configurational freezing'' and related methods~\cite{Owicki1977-ij, Nicolini2009-vl, Chelli2012-yt, Giovannelli2014-rb, Giovannelli2017-dm, Giovannelli2017-tf}, ``locally enhanced sampling'' and related methods~\cite{Roitberg1991-lw,Verkhivker1992-dl}, and solute-focused methods such as ``solvent-solute splitting''~\cite{Leimkuhler2016-ci}, ``replica exchange with solute tempering''~\cite{Liu2005-tt}, and ``alchemically enhanced sampling''~\cite{Lee2023-xo}.
In brief, our work differs from these in that it provides a flexible recipe for extracting subproblems while preserving the original target distribution.
In this work, we have only applied conventional MD to the extracted subproblem, but the approach is expected to be complementary with enhanced sampling methods.

The radial restraint developed in Section \ref{section:radial_selection} is similar in motivation and substance to the ``elastic barrier'' developed in ~\cite{Giovannelli2016-fx}, but is derived differently (so that it can lead to exact sampling without taking the limit of an infinitely stiff restraint).
The Bernoulli augmentation in Section \ref{alg:generic_local_resampling} drew on a computational statistics paper introducing ``Firefly Monte Carlo''~\cite{Maclaurin2014-mn}.

A more detailed discussion of related work is provided in Appendix \ref{appendix:related_work}.

\section{Conclusions}
In this work we have described a flexible approach for extracting lower-dimensional subproblems from high-dimensional sampling problems, specialized the approach for molecular dynamics implementation, and validated it in the context of hydration and binding free energies.

\section{Acknowledgements}
The authors thank:
Paul Hawkins for helpful discussions about measuring sampling bias, and
Charlles Abreu and Ana Silveira for helpful discussions about Hamiltonian Monte Carlo.
The authors thank Gavin Crooks, Peter Eastman, Michael Shirts, and David Minh for insightful feedback on a draft of the manuscript.

The idea in Section \ref{section:local_resampling}-\ref{section:radial_selection} percolated from some long-term influences, esp. from
John Chodera and Marcus Wieder (instilling a rigorous approach to sampling, and working on a project that focused on simulating water droplet subsystems),
Dominic Rufa and Patrick Grinaway (for years of helpful discussions about SMC and all things related to sampling),
and Bas Rustenburg (for presenting the paper Firefly Monte Carlo~\cite{Maclaurin2014-mn} at a journal club in 2016).


\section{Author contributions}
$\dagger$ denotes equal contribution
\begin{itemize}
    \item JF$^\dagger$: methodology (local resampling), prototype implementation, investigation (local SMC, local HMC, local RBFE radius scan), describing relation to prior work, drafting manuscript, code reviews
    \item FY$^\dagger$: methodology (alternating global and local moves), investigation (assessing bias in staged AHFE and staged RBFE), GPU implementation, performance benchmarking, drafting manuscript, code reviews
    \item MW: implementation (nonbonded interaction group), methodology (bias measurement), code reviews, editing manuscript
    \item JK: code reviews, identifying related work, editing manuscript
    \item YZ: supervision, suggestion of test systems, implementation, code reviews, editing manuscript
\end{itemize}

\textbf{ASSOCIATED CONTENT}

Supporting Information available: Detailed related work and extended simulation details.

\printbibliography

\newpage

\appendix

\renewcommand{\thepage}{S-\arabic{page}}
\setcounter{page}{1}

\begin{center}
    \Large \textbf{Supplementary information: A local resampling trick\\for focused molecular dynamics\\}
    \vspace{1em}
    \normalsize Joshua Fass$^\dag$, Forrest York$^\dag$, Matthew Wittmann, Joseph Kaus, Yutong Zhao*\\Computation, Relay Therapeutics, Cambridge, MA, 02139, US\\ *Correspondence: \texttt{yzhao@relaytx.com}
\end{center}

\newrefsection


\renewcommand{\thefigure}{S\arabic{figure}}
\setcounter{figure}{0}

\section{Relation to other methods}
\label{appendix:related_work}

\subsection{Configurational freezing, dynamical freezing, preferential sampling}
\label{appendix:configurational_freezing}

A closely related series of papers by Chelli and co-workers investigated methods for freezing spatially defined regions of a molecular system, in the context of nonequilibrium switching calculations.

In ``dynamical freezing''~\cite{Nicolini2009-vl}, mass-velocity scaling is used to slow down particles outside an interesting region.

In ``configurational freezing''~\cite{Giovannelli2014-rb,Nicolini2011-ty,Chelli2012-yt}, Monte Carlo proposals that move particles into or out of the defined region are rejected.
This can be interpreted as an implementation of the recipe described in Section \ref{section:local_resampling}, in a way that is not amenable to using molecular dynamics to implement the transition kernel $T$ (due to discontinuities in $p_\text{selection}(\zs | \xs)$ as a function of $\xs$ at fixed $\zs$).
An early predecessor of this approach is Owicki's ``preferential sampling''~\cite{Owicki1977-ij}.


In ``elastic barrier dynamical freezing''~\cite{Giovannelli2016-fx}, a heuristic restraint function is added to prevent particles from leaving the mobile region.
This is motivated by a similar picture to the ideal gas example in Figure \ref{fig:naive_conditional_resampling}, adding a restraint to prevent ``outward flow'' from the selected region.
It can be interpreted as an implementation of the recipe described in Section \ref{section:local_resampling}, but where the function $p_\text{selection}$ and the transition kernel $T$ are not fully self-consistent.
In particular, $p_\text{selection}(z_i | \xs)$ is 1 for particles $i$ within the mobile region and 0 for particles outside the region, but the transition kernel $T$ would be consistent with a small probability of selecting particles outside the region.
(They can become self-consistent in the limit of an infinitely stiff restraint function, for example.
The bias introduced by this inconsistency is controlled by the user picking a sufficiently stiff restraint function, and is measured to be small in practice in ~\cite{Giovannelli2016-fx}.
However, these would not be exactly self-consistent even in the limit of an infinitely stiff restraint, if the mobile region is defined based on distance to a specific particle.)

In contrast, the construction in Section \ref{section:local_resampling} can be applied correctly in more general contexts (e.g. with weak restraints, or in principle, with selection probabilities that depend on the configuration in more sophisticated ways).

\subsection{NCMC}
\label{appendix:ncmc}
The main text presented Algorithm \ref{alg:generic_local_resampling} as an auxiliary variable method, but it can also be productively viewed as a special case of nonequilibrium candidate Monte Carlo~\cite{Nilmeier2011-hk} (NCMC).
We briefly review NCMC, although with some loss of generality -- see ~\cite{Nilmeier2011-hk} for a complete description).

An NCMC move involves simulating a path of auxiliary variables $\mathbf{X} = (\xs_0, \xs_1, \dots, \xs_T)$ according to a protocol $\Lambda$, and enforcing a pathwise form of detailed balance on these auxiliary variables.
To define an NCMC method, one must specify: Two conditional distributions\footnote{In ~\cite{Nilmeier2011-hk}, these are presented as a single distribution $p(\Lambda | \xs_0)$, with two behaviors depending on time-reversal.} of auxiliary protocol variables $p_f(\Lambda | \xs_0)$ and $p_r(\Lambda | \xs_T)$, and two tractable conditional distribution of paths $p_f(\mathbf{X} | \xs_0; \Lambda)$, $p_r(\mathbf{X} | \xs_T; \Lambda)$.
Defining the detailed balance condition in terms of these auxiliary variables makes the acceptance probability tractable, even if the marginal proposal probability $p(\xs_T | \xs_0 ; \Lambda) = \int_{\xs_1, \dots, \xs_{T - 1}} p_f(\mathbf{X} | \xs; \Lambda)$ is intractable.
(An analogous approach underlies tractable importance weights in SMC samplers~\cite{Del_Moral2006-ov}.)
The overall NCMC acceptance probability is roughly $(p_\text{target}(\xs_T) p_r(\mathbf{X} | \xs_T; \Lambda) p_r(\Lambda | \xs_T)) / (p_\text{target}(\xs_0) p_f(\mathbf{X} | \xs_0; \Lambda) p_f(\Lambda | \xs_0))$ -- but again this is stylized  -- please refer to original reference~\cite{Nilmeier2011-hk}.

An NCMC protocol $\Lambda$ typically consists of a sequence of ``propagation'' and ``perturbation'' kernels, and a proposal trajectory $(\xs_0, \xs_1, \dots, \xs_T)$ is generated by applying that sequence of steps to an initial state $\xs_0$.
A common special case is for the ``perturbation'' steps to change the energy function, and for ``propagation'' steps to update $\xs$ using MD or MCMC moves, based on the current energy function.

The method in Section \ref{section:local_resampling} can be seen as a special case, where we associate a protocol $\Lambda$ with every possible partition $\zs$ of the system into moving and frozen subsets, so that the conditional distribution $p(\Lambda | \xs)$ corresponds to $p_\text{selection}(\zs | \xs)$.
A given protocol $\Lambda$ instantaneously switches on restraints (performing some work), propagates (in some way assumed to be already accounted for), and then instantaneously switches off the restraints (extracting some work).
If the restraint potential is exactly matched to $p(\Lambda | \xs_0)$ (as in $q_\text{restrained}(\xs)$ of Algorithm \ref{alg:generic_local_resampling}), then the contributions from the protocol selection and the restraint switching can cancel out, leading to unit acceptance probability.
Relaxing this matching aspect of Algorithm \ref{alg:generic_local_resampling} in the framework of NCMC may be fruitful -- although it will reduce the acceptance rate from 1, it may buy some flexibility.

Aside from being a special case of NCMC, Algorithm \ref{alg:generic_local_resampling} is anticipated to be a useful propagation kernel within NCMC, as in ~\cite{Giovannelli2014-rb} (which demonstrated the utility of focused single-particle moves within NCMC).

\subsection{Metropolis-within-Gibbs}
\label{appendix:gibbs}

Metropolis-within-Gibbs can be seen as a special case, where the selection of particles $i$ to update does not depend on the configuration $\xs$: $p_\text{selection}(\zs | \xs) = p_\text{selection}(\zs)$.

There are many variants of Gibbs sampling, such as (1) using random scan order vs. deterministic scan order, (2) single-site vs. blocked selection of the variable indices, and (3) adapting the scan parameters over time based on the history of the chain.

However, to the best of our knowledge, there does not exist a generalization in the literature where the variable index selection is conditioned on the current values of variables.
This generalization is not required for models where interacting variables can be identified solely based on the index (e.g. if $i, j$ share an edge in an Ising model), but it is required in our applications, since interactions will vary depending on the current state (e.g. when the distance between particles $i$, $j$ is below a threshold).



\subsubsection{Constrained HMC within Gibbs}
\label{appendix:constrained_hmc}

In ~\cite{Spiridon2017-fq}, constrained MD was used as a Gibbs sampling move to sample a flexible system.
For example, torsion dynamics could be applied to improve sampling of a system whose bond and angle terms are flexible, or rigid body dynamics could be applied to a system which does not contain any rigid bodies.
This requires careful consideration of the Jacobian of the transformation from Cartesian coordinates to generalized coordinates.
``Generalized equipartition''~\cite{Jain2012-qv} needs to be considered when initializing velocities, simulating the constrained dynamics, and computing acceptance probabilities.

The present work is similar in several respects -- an MD-based sampler is applied to a lower-dimensional subproblem, sampler moves that target subproblems can be alternated with sampler moves that target the full problem, and the choice of subproblem depends on the current state of the system.
If viewed as applying constrained MD, the present work applies more trivial constraints than those considered in \cite{Spiridon2017-fq}, but allows greater flexibility in the selection of subproblems.
In the case of selecting a subset of $K$ particles to move, the Jacobian of the transformation from Cartesian coordinates to ``generalized coordinates'' $\mathcal{X}^N \to \mathcal{X}^{K}$ simply contains a $K$-by-$K$ identity matrix, padded with zeros, and some relevant computations involving that Jacobian become trivial.
For example, a velocity distribution satisfying ``generalized equipartition'' is the standard Maxwell-Boltzmann distribution, restricted to the selected particles.

In ~\cite{Spiridon2017-fq}, the constraint parameters (such as the lengths of constrained bonds) are a deterministic function of the current state of the system, and these parameters are preserved by the constrained dynamics.
In the present work, the selection $\zs$ is a random variable conditioned on the configuration $\xs$.
It is also possible in principle to update $(\xs, \zs)$ jointly, but we did not investigate this possibility.

\subsection{Solvent-solute splitting}
\label{appendix:solvent_solute_splitting}

In works such as \cite{Leimkuhler2016-ci}, atoms are partitioned into two groups of indices -- the solute (of most direct interest), and the solvent (a larger number of environment particles, of lower interest), so that solvent-solvent interactions can be evaluated 
less frequently than solute-solvent interactions.
(As mentioned in Section 5 of \cite{Leimkuhler2016-ci}, the number of solvent-solvent interaction forces is much larger than the number of interactions between the solute and solvent, or internal to the solute, and are the dominant cost in typical MD simulations of biomolecules in detailed solvent.)
This performance optimization is applied in a multiple-timestep (MTS) framework.

In the present work, the decomposition of the system into a small interesting region and a large exterior region can be done on the fly, in a way that depends on the configuration, and embedded in an exact method for canonical sampling.
In contrast to MTS (which updates all particles every step, but reduces the frequency of computing expensive forces), this approach updates a subset of particles.

\subsection{Mixed resolution methods}
\label{appendix:multiresolution}

Some methods with a similar motivation require changing the model definition, e.g. to introduce coarse-grained sites or implicit solvent beyond the explicitly modeled interesting region.
For example, ``active site dynamics'' and related prior methods \cite{Berkowitz1982-bw,Brooks1983-cy,Brooks1985-gh,Beglov1994-oo} extract a small region to focus on, and define a stochastic buffer force on the region's boundary.
Other works in a similar vein include ``semi-explicit'' methods such as ~\cite{Brower1997-ob, Kimura2000-gx}, which surround the interesting region with a sphere of fluctuating point charges to model the reaction field of the bulk, and ``minimalist explicit'' solvent methods such as \cite{White2006-oz}, which restrain a small number of explicit waters near interesting regions.
These works suggest future avenues for defining selection regions.
For example, spherical restraints and nearest-loop-atom restraints were considered in \cite{White2006-oz}.

Another series of recent works with similar motivation~\cite{Wagoner2011-pv,Wagoner2013-yx,Wagoner2018-az} considered a case where the ``interesting sphere'' is treated with explicit solvent, and the exterior of the sphere is treated with implicit solvent.
Explicit solvent particles can enter and exit the sphere via specialized Monte Carlo moves.
The boundary between the explicit region and implicit region is allowed to vary dynamically.

In contrast, we treat the entire system using ``full resolution'' explicit solvent, which can simplify practical implementation.
For example, it presents an alternative to grand-canonical Monte Carlo (with a parameterized chemical potential for particles fluctuating into and out of the explicitly modeled region).
Instead of requiring the modeller to introduce any new functional forms for the effect of the bulk, it carries some implicit assumptions, e.g. that dynamics within the focus region are not too coupled to slow fluctuations of the bulk.

We are also free to change the selection dramatically between moves (focusing on multiple regions, larger or smaller regions, etc.), rather than having to make a single good choice for the partition between interesting subregion and uninteresting surroundings.
This flexibility may be useful in the context of adaptive methods such as SMC (where each copy of the system can propagate a differently sized interesting region), although we did not yet investigate this possibility.
This flexibility also may be appealing for the purposes of incremental adoption -- rather than committing to a single high-stakes choice of the partition between interesting region and bulk for the full duration of a simulation, we can randomize this choice many times in the course of a single simulation.

\subsection{REST}
\label{appendix:rest}

``Replica exchange with solute tempering'' (REST) ~\cite{Liu2005-tt,Wang2011-bp,Wang2012-bn} selectively heats up or softens interactions involving atoms that are selected by index.
Coupled with replica exchange, this allows for improved conformational sampling at the thermodynamic state of interest.

For example, in a binding free energy calculation, the ligand and nearby residues may be selected, and in a relative protein-protein binding calculation the residues in the REST region may be selected based on proximity to a mutated residue~\cite{Zhang2023-ol}.

Variants of REST are effective and widely used for free energy calculations.
Local resampling differs from REST in that
(1) the selection can be based on geometry, not just index (for example, allowing waters to be selected),
(2) deselected particles are frozen, rather than propagated, which can be exploited to reduce the cost per step,
(3) local resampling is agnostic to ``internal'' choices of how to enhance sampling of the selected region,
(4) local resampling does not rely on replica exchange.

\subsection{Firefly Monte Carlo}
\label{appendix:firefly_mc}
While differing in motivation, the ``Firefly Monte Carlo'' method by \cite{Maclaurin2014-mn} is notable for using auxiliary binary random variables to mask out parts of a problem.
Auxiliary binary random variables were used in \cite{Maclaurin2014-mn} to mask out individual terms in an expensive sum, in a way that can be accounted for.
In that work's context of sampling a posterior distribution, these terms correspond to data-dependent likelihood terms.
In the context of molecular simulation, this would be analogous to ``turning off subsets of interactions'' in the energy sum while allowing all particles to move.
By contrast, the current work aims to ``freeze subsets of particles'' while still considering all of their interactions.

\subsection{Locally enhanced sampling}
\label{appendix:locally_enhanced_sampling}

``Locally enhanced sampling'' (LES) ~\cite{Roitberg1991-lw, Simmerling1998-qf} is a method where a subset of the system is selected and $N$ replicas are created from this selection.
The replicas interact with the rest of the system via a potential scaled by $1/N$, but do not interact with each other.
In this way, the replicas are expected to sample different configurations, resulting in enhanced sampling.
This method has also been combined with Replica Exchange to create Local Replica Exchange Molecular Dynamics (LREMD) \cite{Cheng2005-xh}.

Our method differs from LES in that:
(1) LES simulates a system that is not thermodynamically equivalent to the original~\cite{Stultz1998-wc,Zheng1997-wk} (although this can be partially mitigated by adding and tuning restraints~\cite{Ovchinnikov2020-he}),
(2) the selection of atoms in LES is made by index, but our method allows for the selection to depend on the current conformation,
(3) our method can freeze atoms that have not been selected, and 
(4) our method does not rely on simulating multiple copies of the selected atoms.


\section{Simulation details}
\label{appendix:simulation_details}

\subsection{Freezing the central particle}
\label{section:freezing_central_particle}


In implementing the radial specialization of Section \ref{section:radial_selection}, we may wish to avoid computing the contributions of the second group of terms in eq. \ref{eq:fb_restraint} for reasons of correctness and efficiency, which we can accomplish by holding the central particle $x_c$ fixed.
For brevity, we will refer to this group of terms as ``the deselection restraint'' $U_\text{deselected}$, but the interpretation of these terms remains somewhat unintuitive.
Freezing the central particle in this way does not make the move invalid, and it does not make the move ergodic -- both variants can be valid, and neither variant is necessarily ergodic.

It is easy to define a distance-based restraint function $U(r)$ that is compatible with MD -- namely, that it is at least $C^2$ continuous.
But it may require more care to verify that the transformation $U_\text{deselected}(r) = \beta^{-1} \log(1 - \exp(-\beta U(r)))$ is also compatible with MD.
For example, the associated forces may be much larger ($\max_{0 < r < r_\text{max}} |\partial U_\text{deselected} / \partial r | \gg \max_{0 < r < r_\text{max}} |\partial U / \partial r |$), possibly imposing a more severe MD step size restriction than the one imposed by $U$.
It might not be continuous as a function of $r$, and it is not finite when $U(r) = 0$ (which, for any choice of flat-bottom restraint, can mean that the energy is $+\infty$ for a large region of configuration space).

Skipping the deselection restraint also avoids the $O(N - K)$ computational expense of summing over a possibly large number of deselected particles, although this is unlikely to be a limiting factor in any case.

Finally, we do not expect that skipping these terms by holding the central particle fixed will come at a great cost to sampling efficiency in most cases, relative to allowing the central particle to move subject to this restraint.
This is because the second group of terms in eq. \ref{eq:fb_restraint} already prevents $x_c$ from moving in such a way that its distance to any deselected atom becomes lower than $r_0$.
There may be exceptional cases (such as the selection of a sufficiently small ligand floating in a sufficiently large empty fullerene cage, at specific choices of $r_0$), where there can be a practical difference in sampling efficiency between freezing the central particle vs. allowing it to move subject to the deselection restraint.

For completeness, however, we also implemented the deselection restraint, and did not observe that it imposed additional limits on MD stability when used in combination with the flat-bottom quartic restraint potential described in Section \ref{section:radial_selection}.


\subsection{Performance benchmark}
\label{appendix:benchmark_simulation_details}
NVT only simulations were run on an Nvidia A4000 using Timemachine commit \url{https://github.com/proteneer/timemachine/tree/984f48ee76e2364e29cb78819a3fdfc3cdb9b15d}

\subsection{System Setup For Estimating Bias}
\label{appendix:bias_system_setup}

The ligand was parameterized with OpenForcefield 2.0.0~\cite{Boothroyd2022-vu} along with AM1-BCC~\cite{Jakalian2002-jh} charges (with corrections from the OpenFF Recharge project \url{https://github.com/openforcefield/openff-recharge}).
The water was modeled using TIP3P~\cite{Jorgensen1983-hh}.
Each simulation collected 5000 frames, after 50000 steps of NPT to equilibrate the system.
All systems were configured with hydrogen mass repartitioning (HMR) applied.

\subsection{Parameter Filtering Details}
\label{appendix:parameter_filtering}

The goal of filtering was to come up with a protocol for generating a threshold from the global simulations and then applying that threshold to the simulations with local moves to determine whether the bias was acceptable.

The initial pass of filtering was removing any parameters that produced $+\inf$ atom energies during the simulation or crashed.
These parameters were treated as unstable and no further analysis was performed on the combined parameter sets.

Once we had a set of samples that were at least stable, even if perhaps biased, we needed to thin the trajectories down for analysis.
In the case of the global MD samples frames were collected every 500 steps, however in the simulations that alternated local steps they were selected at different intervals.
To determine the thinning, we computed the autocorrelation time of the energies for global samples as well as the local samples with more than 100 global steps.
The decision to only look at local samples with more than 100 global steps came from noticing, during development, that that fewer numbers of steps could lead to instability, slower autocorrelation times and a higher thinning interval.
We integrated the autocorrelation function to compute a thinning interval.
We computed this thinning interval for both the global and local samples, finding that the global samples had much lower thinning intervals than the local samples.
In the end we selected the thinning interval of 69, which represented the local samples thinning interval, and this thinning was applied to all samples in the rest of the analysis.

The Two-sample Kolmogorov–Smirnov test was selected for being non-parameteric which was important as not of all the energy distributions were normal and the KS statistic was used rather than pvalues is not as sensitive to the samples not being independent and identically distributed.

Several approaches were considered when trying to determine which sets of parameters for the local moves were acceptable in their bias.
While we evaluated different approaches, each produced similar parameter sets in the final evaluation.

\begin{enumerate}
    \item Look at the overlap of per atom energy distributions and select thresholds empirically for the KS 2 Sample statistic, which is used as an approximation of bias, when comparing global and local features.
    \item Use Linear Discriminant Analysis to attempt to distinguish between samples and then use the projection to compute a KS statistic that could be used as an estimation of the bias.
    \item Compute the pairwise KS statistics of the global samples and use that as a threshold that can be used when comparing all the global samples to the local samples. \label{enum:pairwise}
\end{enumerate}

The approach that we cover in this paper is \ref{enum:pairwise}, the pairwise KS sample test.

The protocol was to take all the global samples for the global parameters (friction, timestep) and compute the KS statistic on all the per atom energy distributions.
We then computed the pairwise KS statistic between the global samples and the local samples.
These two sets of statistics were then summarized by the mean for each energy distribution, using the global pairwise statistics as the threshold for the local samples.
Doing this initially resulted in no local samples being identified as having 'acceptable bias', which was unexpected after visualizing the overlap of the distributions.
Scaling up the thresholds from the global samples by 100\% resulted in 67\% (779 of 1152 parameter combinations) of the parameters being identified as low bias.
This scaling up of the thresholds was done due to the initial thresholds being too stringent.
Looking through the distributions of the per atom energies of the two sets, gave us confidence that, while having to hand-tune thresholds, sufficiently similar distributions were being replicated with the parameters, as summarized in Figure \ref{fig:filtering_examples}.
Once the 779 parameter sets that had been determined to have acceptable bias, we wanted to determine what parameters would be feasible for use in equilibrium simulations.

As our equilibrium simulations are typically run with a timestep of 2.5 femtoseconds and a friction of 1.0 / picosecond, we focused only on the parameters sets with those values, though we had done the grid search specifically because we weren't sure local moves under these conditions would be acceptable.
Next the parameters were filtered to only include settings where multiple radii were valid, under the premise that if multiple radii were accepted then the parameters were exhibiting lower bias generally.
Afterwards the parameters were filtered for the lowest number of global steps, where there global steps $\geq 100$ due to the earlier detail of small numbers of global steps produced invalid energies, and lowest number of local steps.
This parameter set goes in line with what had been observed in development had concluded, that having at least 100 global steps mixed in produced more stable results.

\subsection{FreeSolv subset selection}
\label{appendix:freesolv_selection}
A subset of FreeSolv was selected by excluding compounds containing carboxylic acids 
as well as molecules that could not be parameterized under the recharge implementation of AM1BCC (\url{https://github.com/openforcefield/openff-recharge}).
The subset consisted of 611 compounds with an experimental range from -25 to 4 kcal/mol.

\subsection{Staged absolute hydration free energy calculations}
\label{appendix:staged_ahfe_details}

The NPT ensemble was simulated by alternating 15 steps of Langevin dynamics (simulated using the BAOAB integrator~\cite{Leimkuhler2013-iz} with a timestep of 2.5 femtoseconds and friction of 1.0 / picosecond) and 1 Monte Carlo Barostat~\cite{Chow1995-fc, Aqvist2004-bj} move.

All the simulations were equilibrated for 200000 MD global steps (0.5 nanoseconds) using global moves only at a timestep of 2.5 femtoseconds before collecting 2000 frames, where each frame consists of 400 steps (1 picosecond).
For the simulations that alternate between global and local steps, frames consisted of 600 steps (1.5 picoseconds) made up of 100 global steps then 500 global steps.
This choice of sampling less frequently was to select parameters that had been determined to introduce little bias in Section \ref{subsection:local_with_unadjusted}.
The AHFE transformation used 16 pre-optimized lambda windows, with the ligand being lowered into solvent using 4D Decoupling~\cite{Rodinger2005-hd}.
Free energy differences between adjacent windows were estimated using BAR~\cite{Bennett1976-un}.


Experiment scripts can be found at \url{https://github.com/proteneer/timemachine/blob/725ad113b02bf4d4fc33c562938ee0ad1b643e9f/examples/absolute_hydration_energy.py}.

\subsubsection{Comparison of local and global staged AHFE predictions for various local MD settings}
\begin{figure}
    \centering
    \includegraphics[width=0.3\columnwidth]{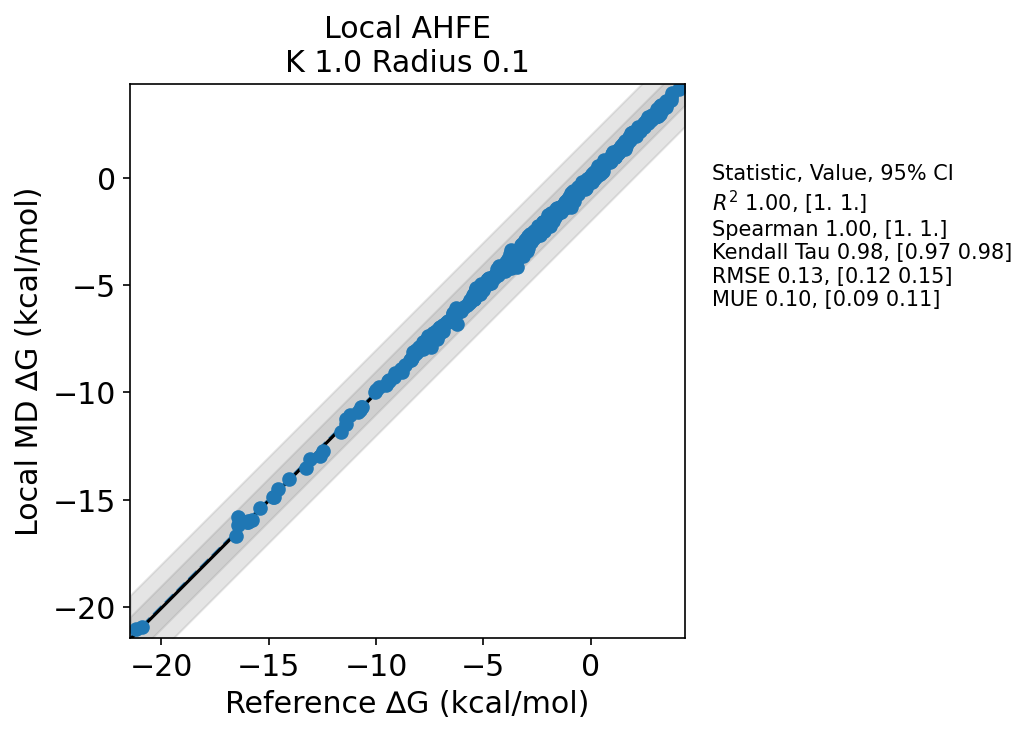}
    \includegraphics[width=0.3\columnwidth]{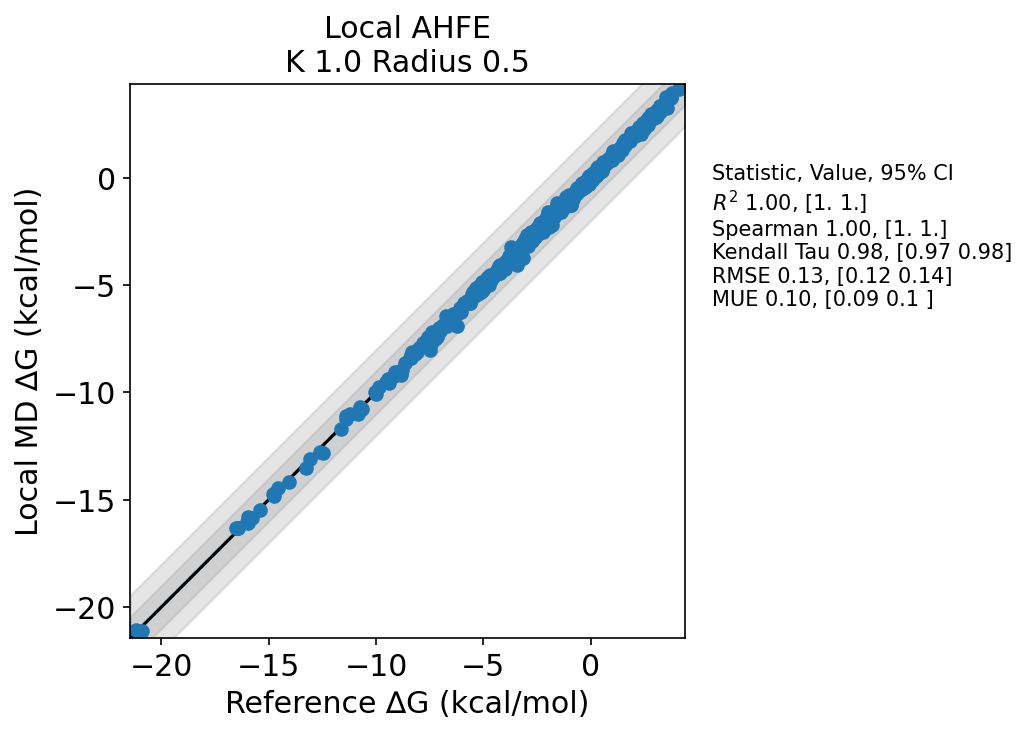}
    \includegraphics[width=0.3\columnwidth]{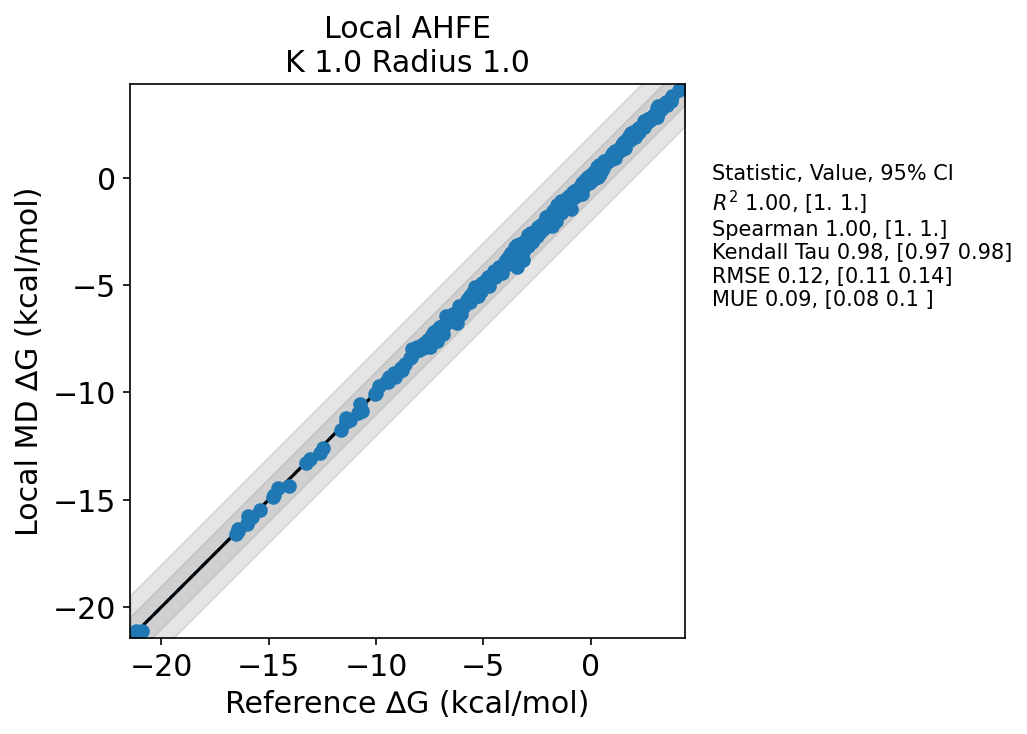}
    \includegraphics[width=0.3\columnwidth]{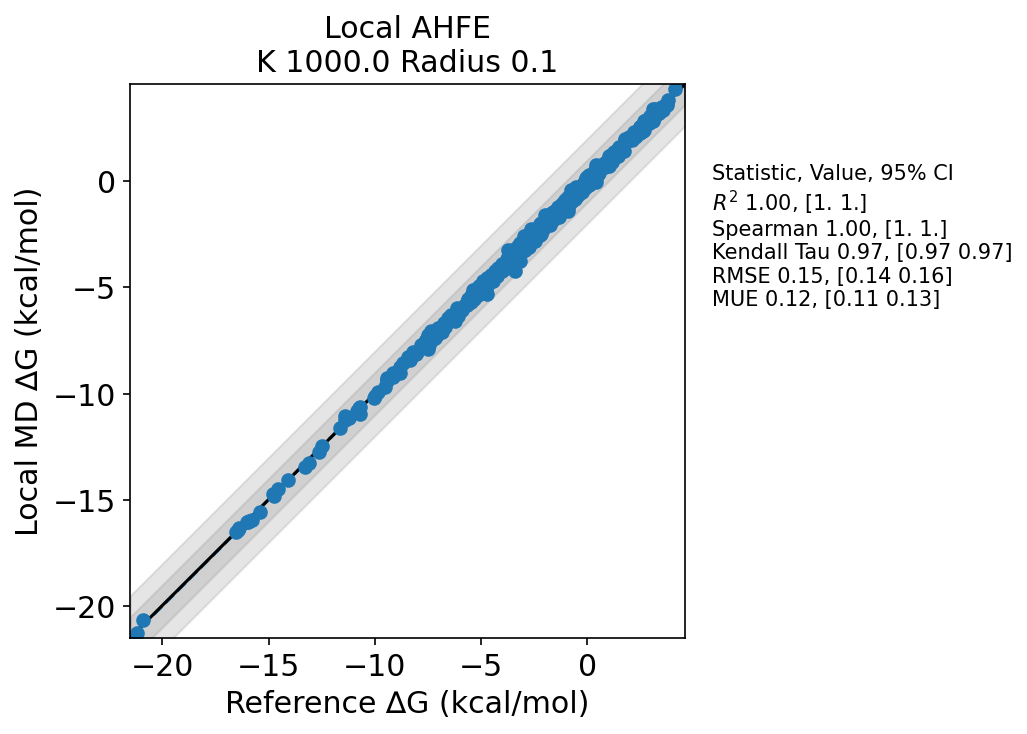}
    \includegraphics[width=0.3\columnwidth]{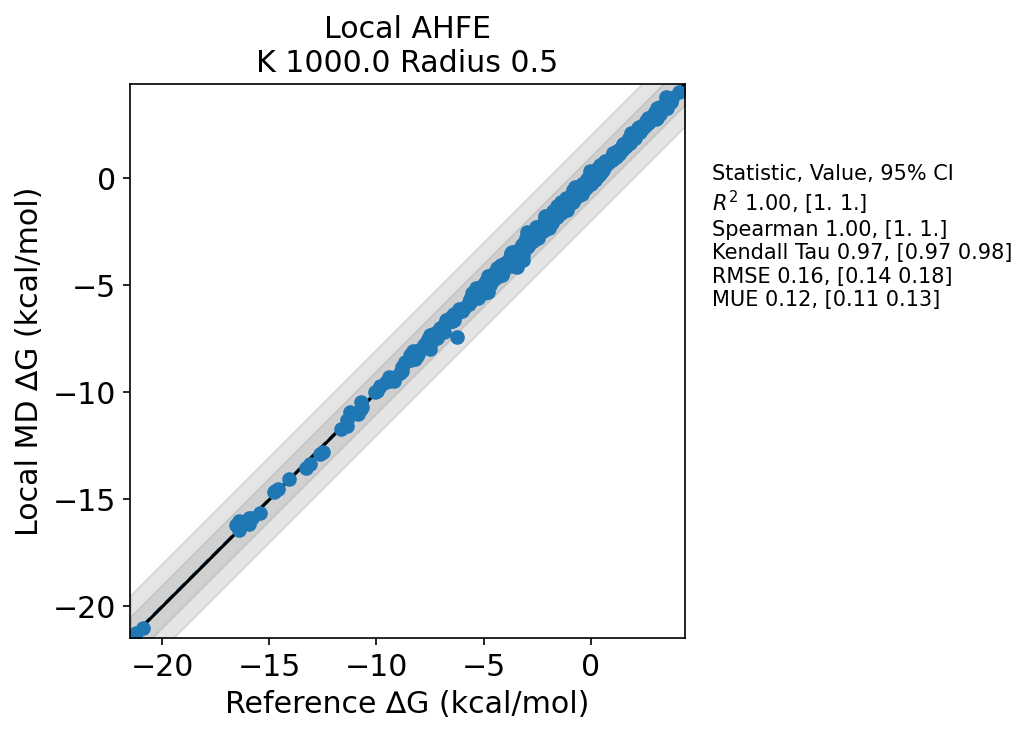}
    \includegraphics[width=0.3\columnwidth]{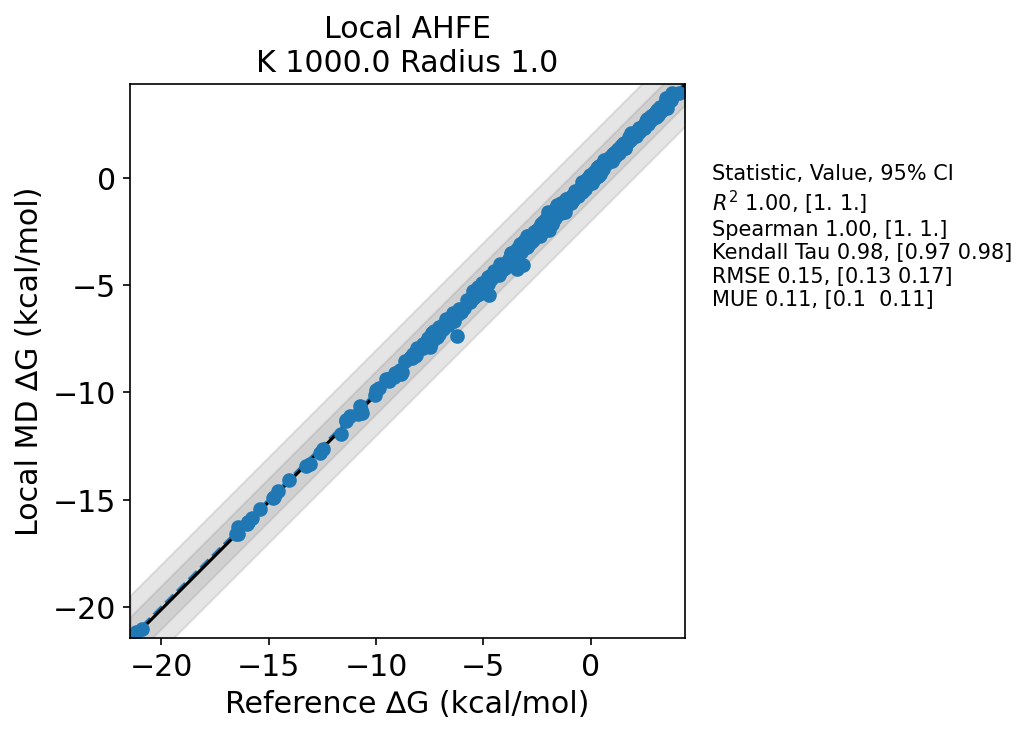}
    \includegraphics[width=0.3\columnwidth]{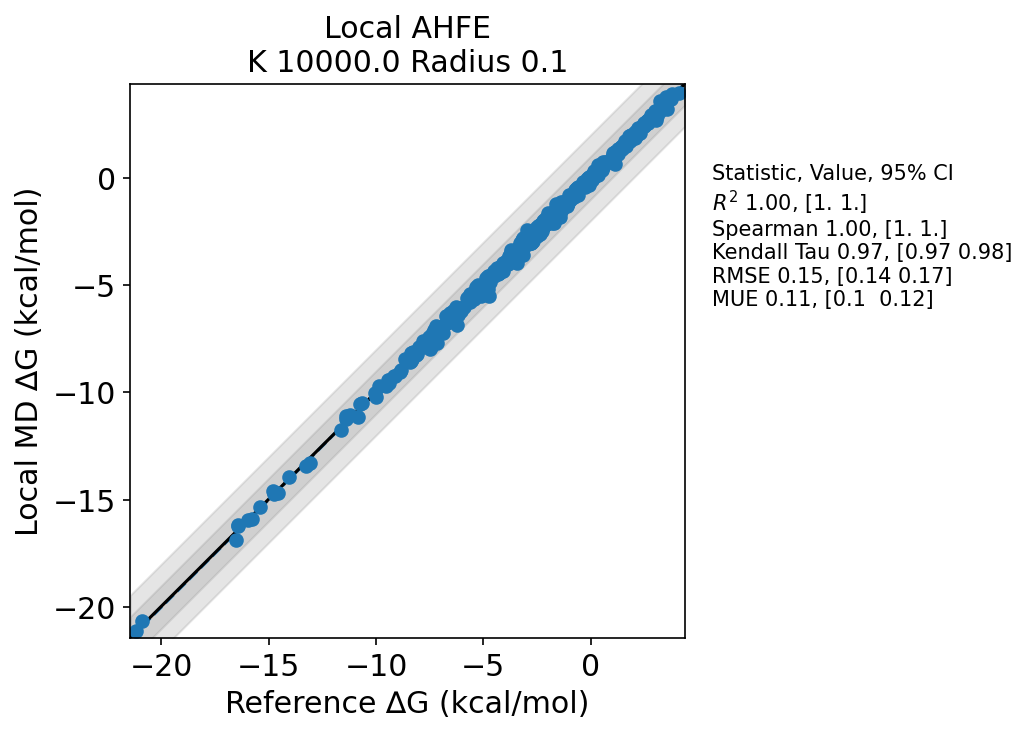}
    \includegraphics[width=0.3\columnwidth]{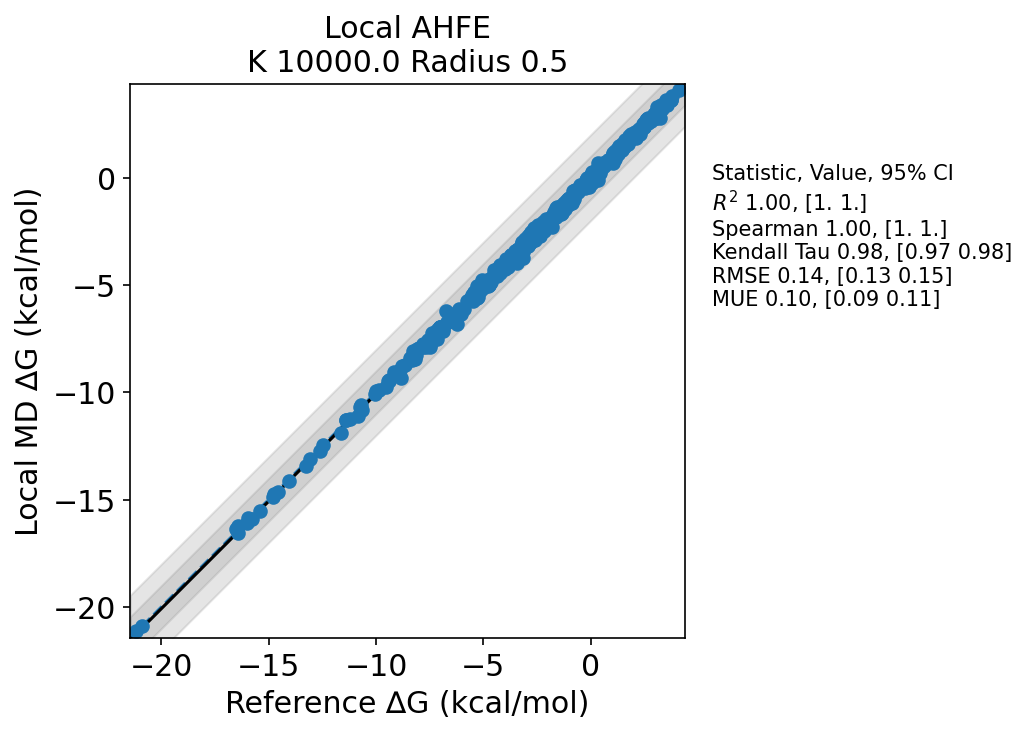}
    \includegraphics[width=0.3\columnwidth]{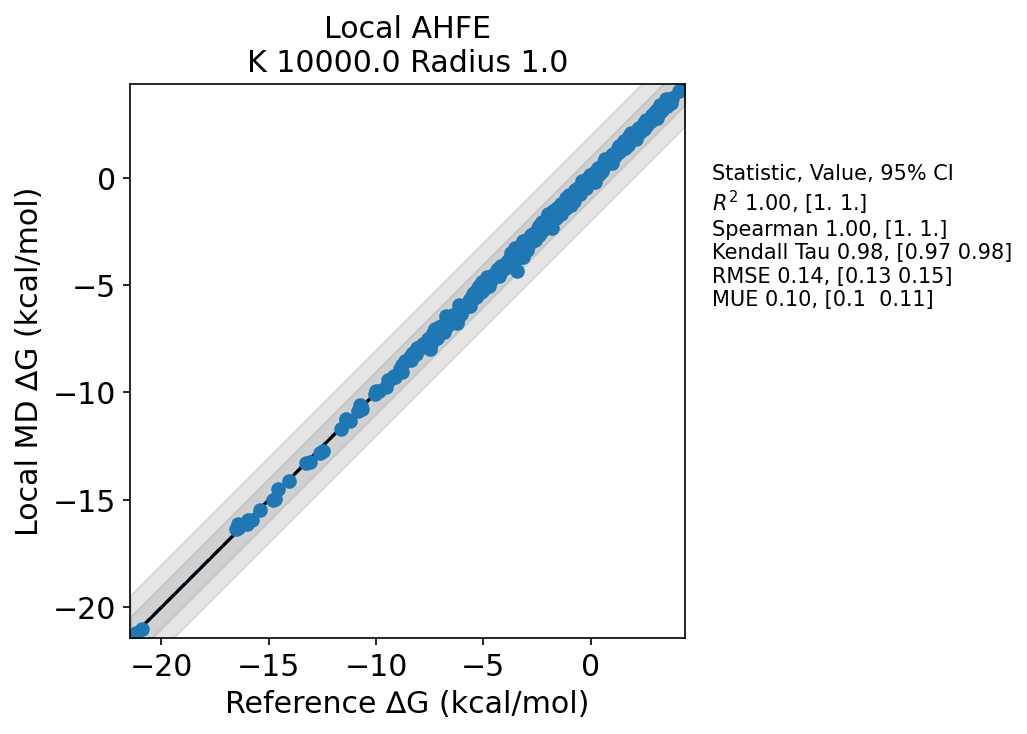}
    \caption{\textbf{AHFE predictions alternating global and local moves compared to global only.}
    Across all values of $k$ and radius, there is good agreement between the reference and local protocols of AHFE.
    Indicating that in the case of an explicit solvent environment these parameters introduce little bias.
    }\label{fig:appendix_local_to_ref_all}
\end{figure}

For the 9 combinations of local MD settings considered in Section \ref{section:staged_ahfe}, we validate that they produce similar AHFE predictions to a reference staged calculation in Figure \ref{fig:appendix_local_to_ref_all}.

\subsubsection{Unconverged Sampling}
\label{appendix:staged_ahfe_slow_sampling}

\begin{figure}
    \centering
    \includegraphics[width=0.3\columnwidth]{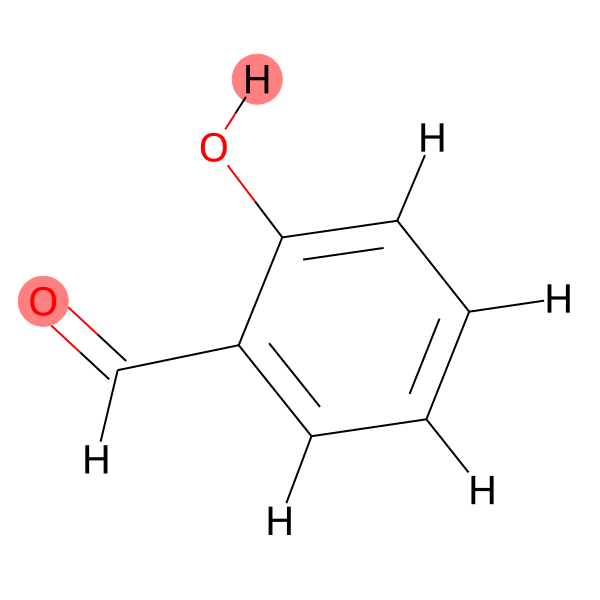}
    \includegraphics[width=0.3\columnwidth]{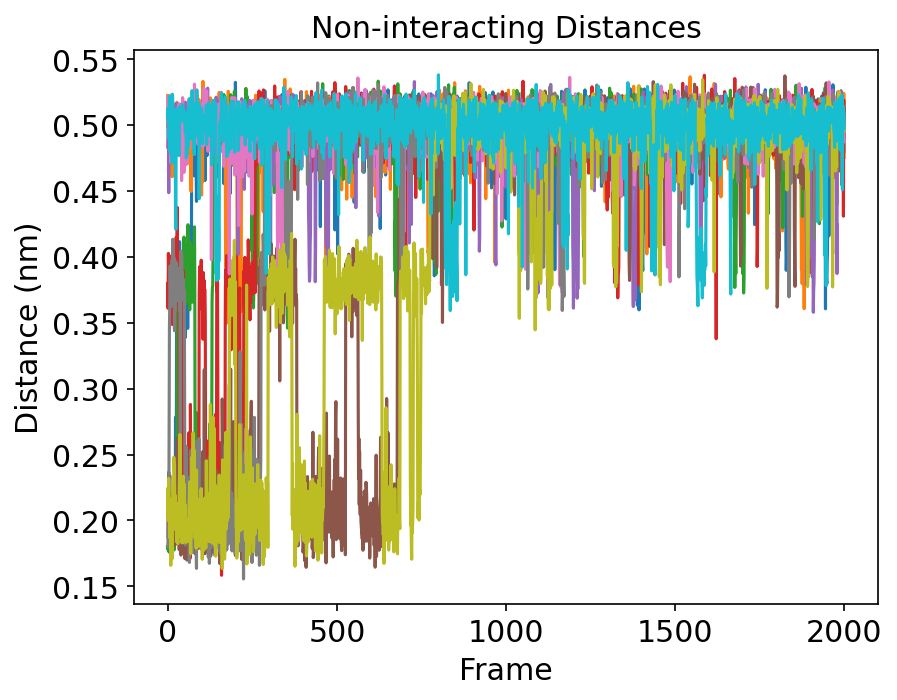}
    \includegraphics[width=0.3\columnwidth]{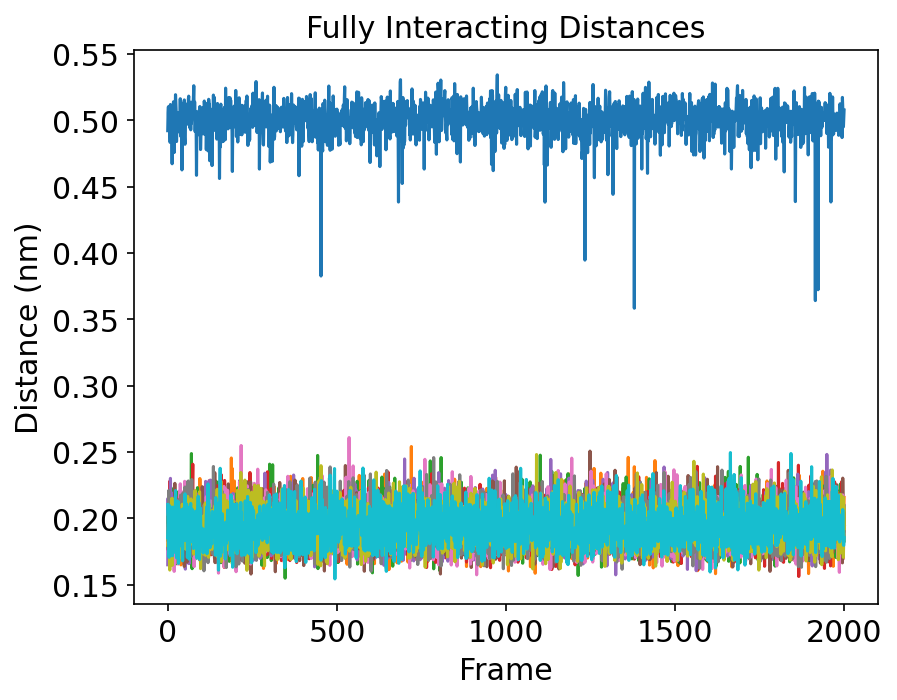}
    \caption{\textbf{Slow sampling of aldehyde and hydroxyl groups in \texttt{mobley\_2850833} across 10 replicates}
    First panel: \texttt{mobley\_2850833} with the atoms of interested highlighted, the subsequent panels compute the distance between these atoms.
    Second panel:  Non-interacting state's distance throughout the simulation.
    Third panel:  Fully interacting state's distance throughout the simulation.
    Without the use of an enhanced sampling method, the sampling of the compound is unconverged resulting in a high variance between replicates.
    }\label{fig:ahfe_slow_sampling}
\end{figure}

In the third panel of Figure \ref{fig:ahfe_comparison} there is an outlier that is greater than 1 kcal/mol in the prediction.
The outlier is compound \texttt{mobley\_2850833} which has unconverged sampling of the aldehyde and hydroxyl groups.
In Figure \ref{fig:ahfe_slow_sampling} the distance between the oxygen of the aldehyde group and the hydrogen of the hydroxyl group is plotted, showing the inconsistent sampling.
It is this unconverged sampling that results in the higher variance seen for this compound between global and local runs of AHFE. 

\subsection{Adaptive sequential Monte Carlo calculations}
\label{appendix:smc_details}

Recall~\cite{Del_Moral2006-ov, Dai2022-ke} that in sequential Monte Carlo, a population of walkers 
is first drawn from an initial distribution $p_0$, and then gradually updated (and importance-weighted) using a sequence of transition kernels.
These transition kernels are often chosen to be MCMC kernels targeting a sequence of distributions $(p_t)_{t=0}^T$, in which case the importance weights are updated by $p_t(x)/p_{t-1}(x)$ (but there is greater flexibility~\cite{Del_Moral2006-ov} in the SMC framework). 
In applications of SMC to molecular simulation, the MCMC kernels are often chosen to be approximate (i.e. unadjusted MD moves), introducing uncontrolled bias.
The bias may be small in practice~\cite{Suruzhon2022-gt, Gill2018-ag} if the sequence of transitions is gradual enough, and if the approximate MCMC kernels are fast-mixing enough.
In the context of an absolute hydration calculation, $p_0$ may correspond to a thermodynamic state where the solvent and solute are non-interacting, and the sequence $p_t$ may gradually introduce their interactions.

The sequence can be chosen in advance, or it can be determined adaptively.
In particular, we use an adaptive sequential Monte Carlo method that selects the next $\lambda$ window based on the ``conditional effective sample size.''~\cite{Everitt2020-do}.
The number of $\lambda$ windows selected by this method will increase both as a function of transformation difficulty and as a function of how poorly mixing the ``propagation'' steps are.

\subsubsection{Algorithm settings}
Global MD was configured similarly to other investigations in this paper: timestep = 2.5fs, friction = 1.0 / picosecond, HMR applied.
However, the Barostat was run every 5 steps, which is more frequent than in other sections.
Initial samples were generated using a burn-in of 50,000 steps and a thinning interval of 1000 steps.

Local MD was configured similarly, with a radius = 1 nm, $k$ = 1000.0, num steps = 250.

Sequential Monte Carlo was configured with the number of particles = 100, beta = 0.95, resample threshold = 0.5, resampling method = multinomial.


\subsubsection{Hardware}
Timings were obtained on a cluster of 10 NVIDIA GeForce 2080 Ti GPUs.



\subsubsection{Neglected adaptation bias}

Using the same particles for both adaptation and estimation in SMC can introduce an estimation bias.
This bias can be removed by performing the adaptation using one set of particles, and using a different set of particles for estimation, but this often appears unnecessary in practice~\cite{Suruzhon2022-gt}, and was neglected in this application.
However, in cases where the adaptation is much more flexible~\cite{Arbel2021-yl, Rizzi2021-rd} than the current adaptation of selecting a single scalar $\lambda$ increment, de-biasing may be more important.

\subsubsection{Future performance optimizations}
It is also possible that future implementations may achieve greater speed-up by:
(1) tuning other SMC settings such as number of particles (here chosen arbitrarily as 100),
(2) exploiting the additional parallelism exposed in the SMC application (i.e. running the 100 sampling subproblems in parallel within each window, rather than sequentially),
(3) reducing the cost of energy evaluations during adaptation (exploiting the fact that computing $U(\xs, \lambda)$ at fixed $\xs$ but variable $\lambda$ can be cheaper than computing $U(\xs, \lambda)$,
(4) further improvements in time per local MD step,
(5) using a pre-computed reservoir of water box samples (generating water box samples is a source of repeated work across these calculations).

\subsection{Staged relative binding free energy calculations}
\label{appendix:rbfe}
The edges selected for the transformations was the same edges as provided in the SI of \cite{Wang2015-dx}.

For the solvent leg of the simulations each ligand was placed into a 4 nanometer cubic box containing 2094 TIP3P waters. For the complex leg the ligand was placed into a cubic box containing the Tyk2 complex with padding of 1 nanometer of water for 5714 TIP3P waters and 4670 protein atoms for the complex leg.
The systems were parameterized the same way as described in Section \ref{appendix:staged_ahfe_details} with the addition of  Amber ff99SB-ILDN forcefield~\cite{Lindorff-Larsen2010-fu} for the protein atoms.
The reference NPT ensemble was simulated by alternating 15 steps of Langevin dynamics (simulated using the BAOAB integrator~\cite{Leimkuhler2013-iz} with a timestep of 2.5 femtoseconds and friction of 1.0 / picosecond) and 1 Monte Carlo Barostat~\cite{Chow1995-fc, Aqvist2004-bj} move.
In the local NPT ensemble, the Monte Carlo Barostat moves were only made every 15 global steps and disabled during local moves.
All simulations were equilibrated for 200000 global MD steps (0.5 nanoseconds) before collecting 2000 frames, every frame containing 400 steps (1 picosecond).
In the case of the simulations with local moves, frames were also collected every 400 steps, only with 100 global steps then 300 local steps run before collecting a frame.
In the staged AHFE simulations setup detailed in Section \ref{appendix:staged_ahfe_details} the parameters used were different to using pre-validated parameters.
This difference was chosen due to the lack of sensitivity to local MD parameters that was seen in Figures \ref{fig:ahfe_comparison} and \ref{fig:local_ahfe_errors} and to more closely mirror the reference protocol.
The RBFE transformation was performed across 48 lambda windows using bisection to determine the windows.
For bisection, each window was sequentially added between the neighboring windows with the highest BAR error.

The script used to run the RBFE calculations can be found at \url{https://github.com/proteneer/timemachine/blob/3e13b430cf8671728567fec5c03a4329ba79e1ea/examples/relative_binding_free_energy.py}.

Simulations were run on Nvidia A10 GPUs, with the reference edges taking ~16 hours to complete.
The simulations which used local moves showed a 10\% speed up, with only ~14 hours per edge to complete.

\section{Neighborlist Modifications}

For the Timemachine MD engine we were able to implement this fast localized MD by modifying our neighborlist to act on disjoint sets of particles, adding a new nonbonded potential and making our nonbonded potential more flexible.
To ensure correctness every free particle must interact with the frozen particles as the free move in the simulation.
We modified our neighborlist to be able to compute interactions between the free particles and the frozen.
This reduces the cost of each neighborlist construction between frozen and free as the free particles are often much fewer than the total system and much of the frozen can be cheaply skipped during bounding box checks.
With the addition of the new neighborlist functionality, we implemented a Nonbonded Interaction Group potential (inspired by OpenMM's~\cite{Eastman2017-uf} InteractionGroup feature) that computes the nonbonded interactions of only a subset of the system against the rest of the system.
This interaction group potential is used to compute the forces of the frozen on the free particles.
The primary nonbonded potential was then modified so that it could compute forces on a subset of the system.
In the case of localized MD we modify the standard nonbonded potential to compute the interactions between all free particles, reducing the size of the neighborlist required to construct.
All the other potentials were left to run in their entirety due to the linear cost and negligible performance benefit relative to the complexity of disabling parts of the potentials.







\section{Local MCMC}
\label{appendix:local_mcmc}

We consider using HMC~\cite{Neal2012-ya} as the transition kernel in Algorithm \ref{alg:generic_local_resampling}.
How does varying the radius of the local region affect the performance characteristics of HMC?

Reducing the radius will move fewer particles, but with a possibly higher acceptance rate and possibly lower cost per move.
The performance of HMC worsens with problem dimension, but at a slower rate than some black-box MCMC moves.
In an idealized case ($N$ independent Gaussians), to maintain a constant acceptance rate as we increase the dimension $N$, we need to decrease the step size as $N^{-1/4}$.




Test system: DHFR benchmark (23558 atoms total). 
Each local move selects a random atom index on the protein, selects a random subset of particles near this index, forms the appropriate restraint, simulates Hamiltonian dynamics for $T = 100$ leapfrog steps, and can be accepted with a probability that depends on the total energy change (including the restraint).
Experiment and plotting scripts are available at: \url{https://github.com/proteneer/timemachine/commit/a00acc01fcdd8849f7fd175bb843db9e281c0765}.

\begin{figure}
    \centering
    \includegraphics[width=0.45\columnwidth]{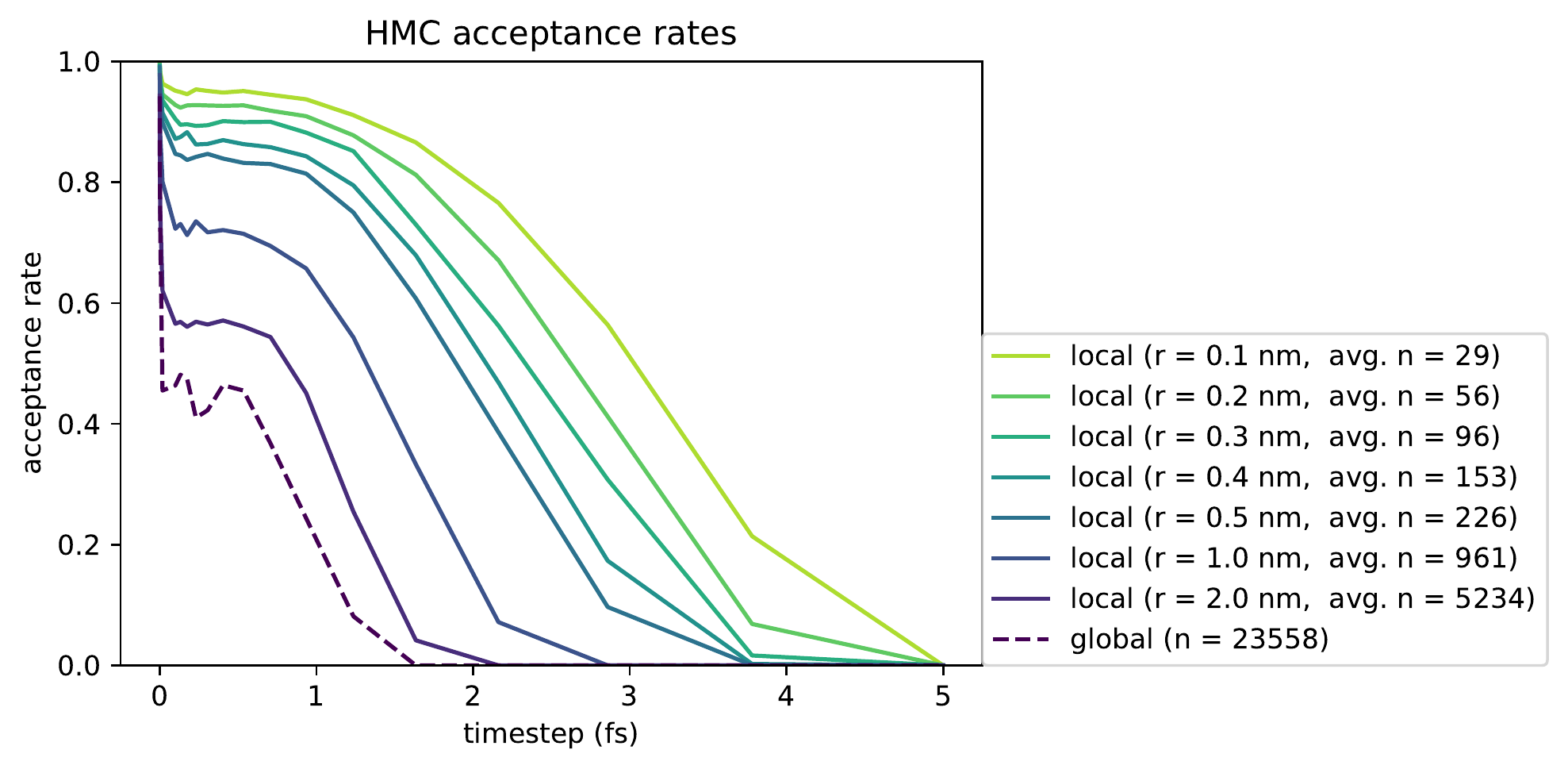}
    \includegraphics[width=0.45\columnwidth]{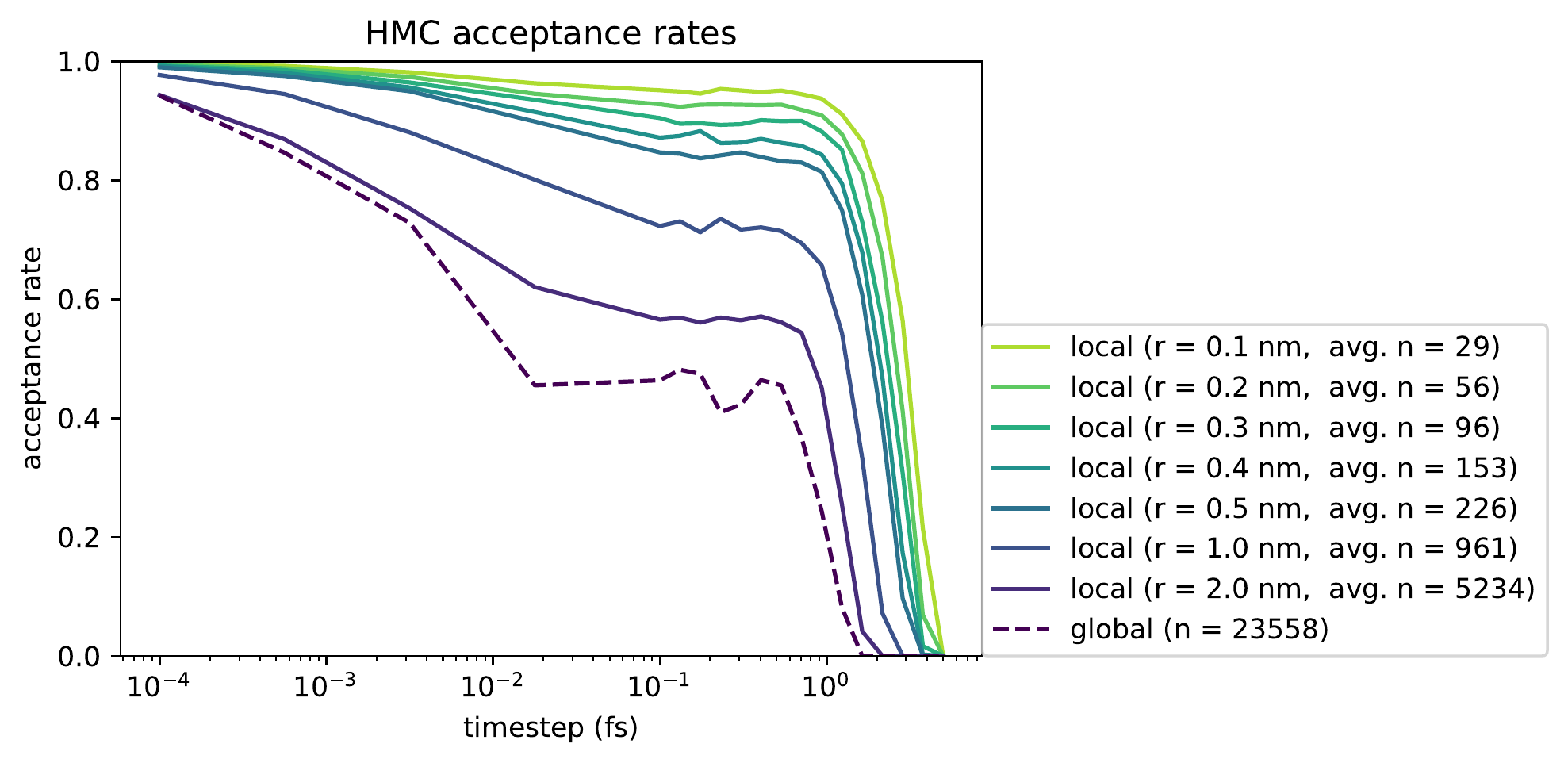}
     \caption{\textbf{Acceptance rate of local Hamiltonian Monte Carlo (HMC) on DHFR test system.}\\
     The dashed line depicts HMC applied to the entire DHFR test system.
     Each solid curve depicts HMC applied to a subset of atoms, based on a different flat-bottom radius, $r_0 =$ [0.1, 0.2, 0.3, 0.4, 0.5, 1.0, 2.0] nm, with the dashed line depicting global HMC.
     The x-axis is the timestep (between $10^{-4}$ fs and 5 fs), and the y-axis is the resulting acceptance rate.
     The number of steps per proposal is fixed at $T=100$.
     The left and right panels are different depictions of the same data.
     First panel: linear x-scale.
     Second panel: log x-scale.
     }
    \label{fig:hmc}
\end{figure}

Our measurements on this system are shown in Figure \ref{fig:hmc}, and these are roughly consistent with theoretical expectations.

\printbibliography

\newpage

\renewcommand{\thepage}{TOC-\arabic{page}}
\setcounter{page}{1}
\renewcommand{\thefigure}{}
\setcounter{figure}{0}

\begin{figure}
\centering
\includegraphics[width=0.9\columnwidth]{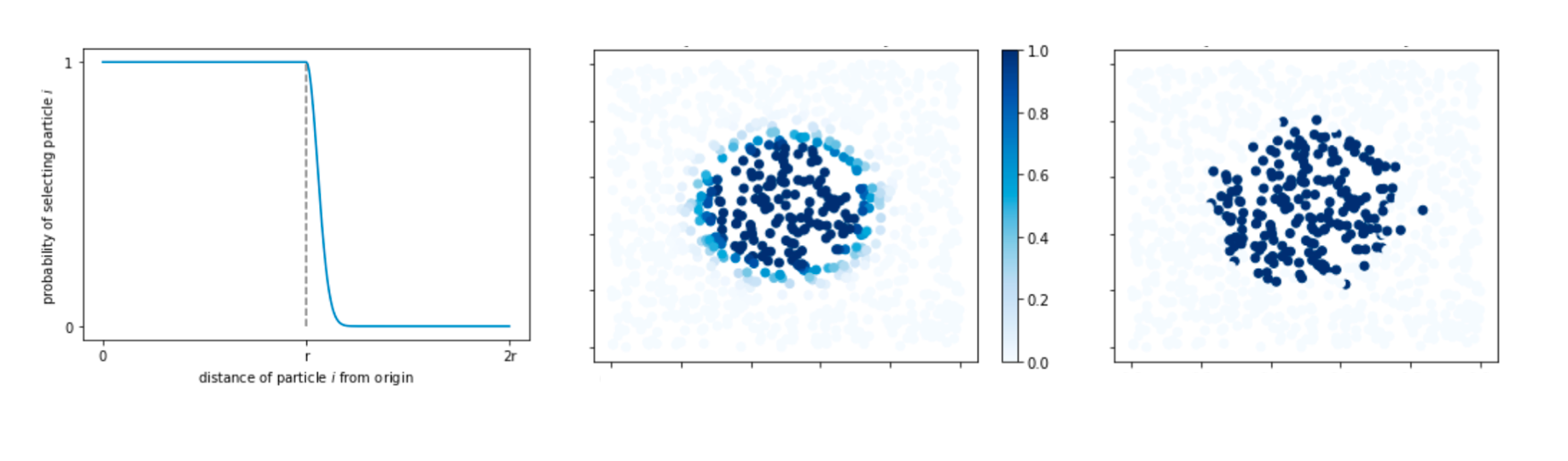}
\caption{\textbf{FOR TABLE OF CONTENTS ONLY}}
\end{figure}

\end{document}